\documentclass{article}
\usepackage{graphicx} 
\usepackage{amsmath} 
\usepackage{amsfonts} 
\usepackage{subfigure}

\begin{document}

\title{Collective states in social systems with interacting \\
learning agents}
\author{Viktoriya Semeshenko\raisebox{0.2cm}{\small (1)} 
\and Mirta B. Gordon\raisebox{0.2cm}{\small (1)}\footnote{corresponding author: mirta.gordon@imag.fr}
\and Jean-Pierre Nadal\raisebox{0.2cm}{\small (2,3)} }
\vspace{2cm}
\date{{\small
(1) Laboratoire TIMC-IMAG (UMR 5525), CNRS - Universit\'e Grenoble 1\\
(2) Centre d'Analyse et Math\'ematique Sociales (CAMS, UMR 8557 CNRS-EHESS), \\
Ecole des Hautes Etudes en Sciences Sociales, Paris \\
(3) Laboratoire de Physique Statistique (LPS, UMR 8550 CNRS-ENS-Paris 6-Paris 7), \\ 
Ecole Normale Sup\'erieure, Paris\\
$\;$ \\$\;$} 
April 16, 2007 \\}

\maketitle

\begin{abstract}
We consider a social system of interacting heterogeneous agents with 
learning abilities, a model close to Random Field Ising Models, where the 
random field corresponds to the idiosyncratic willingness to pay. Given a 
fixed price, agents decide repeatedly whether to buy or not a unit of a 
good, so as to maximize their expected utilities. We show that the 
equilibrium reached by the system depends on the nature of the information 
agents use to estimate their expected utilities.
\end{abstract}

\section{Introduction}
\label{sec:intro}
Individual decisions in social systems are frequently 
influenced by the behaviors or choices of other individuals. 
Besides the obvious case of fashion \cite{NakayamaNakamura}, 
many situations of social influence have been considered 
and analyzed in the literature. They range from sociological 
issues like the decision of attending a bar that may be 
crowded \cite{Arthur94}, a seminar that may have vanishing 
attendance \cite{Schelling}, choosing a movie or a 
restaurant \cite{Becker91}, committing crime \cite{GlaeSaceSche}, 
to political issues such as the decision of joining a riot \cite
{Granovetter78}, voting for or against a new constitution \cite{GaGeSh}, 
etc. 

The first models, proposed by Schelling \cite{Schelling71}, were 
aimed at demonstrating that the collective outcomes when 
individuals interact socially with each other may seem paradoxical 
-- that is, intuitively inconsistent with the intentions of 
the individuals who generate them. In fact, the collective 
states that result from the aggregation of individual decisions 
not voluntarily coordinated, cannot be predicted by any simple 
counting or extrapolation of the individual preferences. 
Schelling \cite{Schelling} built simple models of social 
paradoxes, like the existence of racial segregation in urban 
neighbourhoods despite the non-racist character of the inhabitants, 
the death of a weekly seminar by lack of participants despite their 
interest on it, etc. 
The reason of these paradoxes is to be found in the fact that 
systems with interacting individuals may present multiple 
equilibria. These may be analyzed in a natural way in the 
framework of statistical physics. 

Models of interacting agents facing binary decision 
problems have been considered within an economical 
framework (\cite{Blume93,Orlean95,Durlauf97,GlaeserScheinkman}) 
after Follmer \cite{Fol74} first used the finite-temperature 
Ising model in a homogeneous external field to analyze 
equilibria in a two-goods market. 

In this paper we consider a general model introduced in Gordon 
\emph{et al.} \cite{GoNaPhVa05} and Nadal \emph{et al.} 
\cite{NaPhGoVa06}, where the interacting agents 
have different private willingnesses to pay, i.e. different local fields. 
The individual utilities are the sum of the private and the interactions 
terms. Interactions are assumed to be global and positive, so that 
utilities increase proportionally to the total fraction of buyers. Global 
interactions are pertinent when the individual utilities depend on 
decisions of remote and probably unknown individuals. This is the case of 
the subscription to a telephone network \cite{Rohlfs74,Rohlfs01}, or the 
choice of a standard \cite{MiKa86}, where making the same decision as the 
majority carries advantages. Notice that this kind of aggregated data may 
be easily available through public information. In statistical mechanics, 
this model belongs to the class of mean-field ferromagnetic Random Field 
Ising Models.

As shown by Gordon \emph{et al.} \cite{GoNaPhSe07} when the social 
interactions are strong enough, the system presents multiple (Nash) 
zero temperature equilibria. The one that  is individually and globally 
optimal is called Pareto-dominant equilibrium in economics. 
However, in contrast with physical 
systems where energy and entropy determine the 
actual thermodynamic equilibrium, none of the possible 
equilibria may be ruled out in social systems. 
Multiple equilibria bring on coordination dilemmas 
to the agents. The equilibrium actually reached by the system
depends on the decision making dynamics.

In game theory, mostly limited to two-player games, 
it is usually assumed that individuals possess the skills and 
the information necessary to analyze the 
consequences of all the possible outcomes. Thus, they are 
able to find which is the optimal decision, and thus realize 
the Pareto-optimal equilibrium. 
However, in situations with large numbers of participants 
like the one considered here, or in situations of 
uncertainty, individuals may be unable to grasp 
the information necessary for coordination. 
In fact, they are more likely to rely on beliefs rather 
than on a perfectly rational reasoning to make decisions. Deviations from 
rationality may arise not only in situations of limited 
or incomplete information, but also due to human errors, 
different psychological attitudes with respect to risk, etc. 

We are interested in situations where agents make their decisions 
repeatedly. In that case they may modify their beliefs by learning through 
past experiences. To this end we assume that each agent associates an 
expected surplus or payoff to buying. Once the decisions are made 
according to these expectations, the latter are in turn updated based on 
the grasped information, using a learning rule. This process is called 
learning upon experience in the literature. Behavioral learning is actually 
the subject of important theoretical studies in different disciplines, in 
particular in the context of game theory (see e.g. 
\cite{Benaim_Hirsch_99,Kryazhimskii_etal_00,LaslierTopolWalliser01,Young02})
and in 'econophysics' (see e.g. 
\cite{MinorityGame_book,Marsili_etal_03,Andrecut_Ali_01,NaWeChKi}).
Quite importantly, an increasing access to empirical data allows to compare 
theoretical predictions with observed behaviors 
\cite{Camerer03,WeisbuchKirman}. 

We have studied the equilibria reached by the system for different learning 
rules proposed in the literature to explain outcomes in experimental 
economics. Following  Camerer \cite{Camerer03}, we introduce a small number 
of parameters allowing to study all these rules within a single framework. 
Here we report our most interesting results, obtained through weighted 
belief learning and reinforcement learning. In these settings, buyers 
update their expectations according to their obtained surplus, while 
non-buyers use a degraded information. We compare results for two different 
information conditions. In one of them, that we call $s$-learning, the 
agents estimate the expected surpluses based on actual payoffs. In the 
other, called $\eta$-learning, they estimate them based on the fraction of 
individuals they expect will buy. In the latter case, agents are assumed to 
know the additive structure of the utility function. Our analysis is 
limited to populations of homogeneous learners: all the agents use the same 
learning rule, although they make different initial guesses. Results 
obtained with the standard iterative steepest ascent used to determine the 
local equilibria in Ising model simulations, where at each time step 
individuals make the best decision conditionally to the previous period 
outcome -- a dynamics called myopic fictitious play in game theory --, 
serve as reference for our analysis.

We show that coordination of learners on the optimal (Nash) equilibrium, 
not only in the presence of multiple equilibria but even when the 
equilibrium is unique, is far from being the norm. In 
fact, fairly restrictive conditions are needed. The emerging collective 
state depends strongly on the values of the learning parameters, and is 
very sensitive to the agents' initial beliefs. There are significant 
differences in the aggregate values obtained trough the simulation of both 
learning scenarios. Previous results corresponding to $s$-learning have 
been reported elsewhere \cite{SeGoNaPh06}. The performances along the 
learning paths as well as the incidence of different initial conditions 
on the collective behaviour were thoroughly detailed. In a forthcoming paper
\cite{NaGoSe07} we will present an analytical study of the stationary 
regime attained through the learning dynamics and probabilistic decision-making.

The paper is organized as follows: in section \ref{sec:model}, we present 
the agents model and its statistical mechanics equilibrium properties. In 
section \ref{sec:learning dynamics}, we describe the learning scenarios. 
We present the general settings of our simulations in section \ref
{sec:setting} and the results in section \ref{sec:simulations}. Section 
\ref{sec:conclu} concludes the paper.

\section{Model with heterogeneous interacting agents}
\label{sec:model}
We consider a social system of $N$ heterogeneous agents $(i=1,2,\dots,N)$ 
that must decide either to buy ($\omega_i=1$) or not ($\omega_i=0$) one 
unit of a single good at an exogenous price $P$. Following Nadal \emph{et al.} \cite
{NaPhGoVa06} we assume that each agent $i$ has a willingness to pay $H_i$, 
which represents the maximal amount he is ready to pay for the good in the 
absence of social interactions. The values $H_i$ are assumed to be 
randomly distributed among the agents according to a probability density 
function of average $H$ and variance $\sigma_H$. In addition to this 
idiosyncratic term, the decisions of other agents exert an additive social 
influence on each individual $i$, increasing his willingness to pay if 
others buy. This influence is assumed to be proportional to the fraction of 
buyers (other than $i$):
\begin{equation}
\label{eq.eta_i}
	\eta_i \equiv \frac{1}{N-1}\sum_{k \neq i}^N \omega_k.
\end{equation}
The utility of buying, the surplus, for individual $i$ is:
\begin{equation}
\label{eq.surplus}
	S_i=H_i+ J \eta_i-P,
\end{equation}
where $J > 0$, the weight of the social influence, is assumed to be the 
same for all the agents. 

The equilibrium properties of this model have been analyzed using the mean 
field approximation, for different distributions of the $H_i$ \cite
{GoNaPhVa05,PhaSe07}. More recently, the properties for very general 
distributions have been determined \cite{GoNaPhSe07}. Hereafter 
we briefly summarize the main results, that we illustrate for the 
particular case of the triangular distribution considered in our 
simulations. The latter allows for a complete analytical equilibrium study 
\cite{PhaSe07}. 

In the thermodynamic limit $N\to\infty$, $\eta_i$ in (\ref{eq.eta_i}) and 
(\ref{eq.surplus}) may be approximated by 
\begin{equation}
	\eta\equiv\frac{1}{N}\sum_{k=1}^{N}\omega_k.
\end{equation}
It is useful to introduce the following reduced parameters
\begin{equation}
\label{eq.normalized}
x_i\equiv\frac{H_i-H}{\sigma_H},\,\, j\equiv\frac{J}{\sigma_H},\,\, 
\delta \equiv\frac{H-P}{\sigma_H},
\end{equation}
where $\delta$ is the (reduced) gap between the average willingness to pay 
and the price. 
$x_i$ represents the (reduced) idiosyncratic preference of agent $i$. 
It is a random variable of zero mean and unitary variance, distributed 
among the population according to a probability density function (pdf) 
$f(x)$. 

A rational agent chooses the strategy $\omega_i$ that maximizes his 
(reduced) surplus
\begin{equation}
\label{eq.nsurplus}
	s_i=\delta+x_i+j\,\eta,
\end{equation}
that is:
\begin{equation}
\label{eq.max}
	\omega_i= \arg \max_{\omega \in \{0,1\}} \omega s_i.
\end{equation}
In the thermodynamic limit, the fraction of buyers at equilibrium is equal 
to the probability of buying of the population:
\begin{equation}
\label{eq.etaselfconsistent}
	\eta=\mathcal{P}(\delta+x_i+j\eta>0)=\mathcal{P}(x_i>-s),
\end{equation}
where $s\equiv \delta+j\eta$ is the population's average surplus. 
Solutions to equation (\ref{eq.etaselfconsistent}) give the fractions of 
buyers at equilibrium, which depend on the parameters of the model. The 
properties of the system may be summarized on a \emph{phase diagram}, in 
which the lines separating different regimes of Nash equilibria are 
plotted in the space of the model parameters, namely $j$ and $\delta$. The 
main results are that if the (reduced) strength of the social interactions 
$j$ is larger than a distribution dependent value $j_B=1/f_{max}$ (
$f_{max}$ is the maximum of the pdf), there is a range of values of 
$\delta$ where two different (stable) equilibria coexist: one with a large 
fraction of buyers, the efficient Pareto-optimal one where coordination is 
achieved, and another with a smaller fraction of buyers. This multiplicity 
is a generic property of models with social interactions \cite{GoNaPhSe07}. 

\begin{figure}[ht!]
\centering     
\includegraphics[width=8cm]{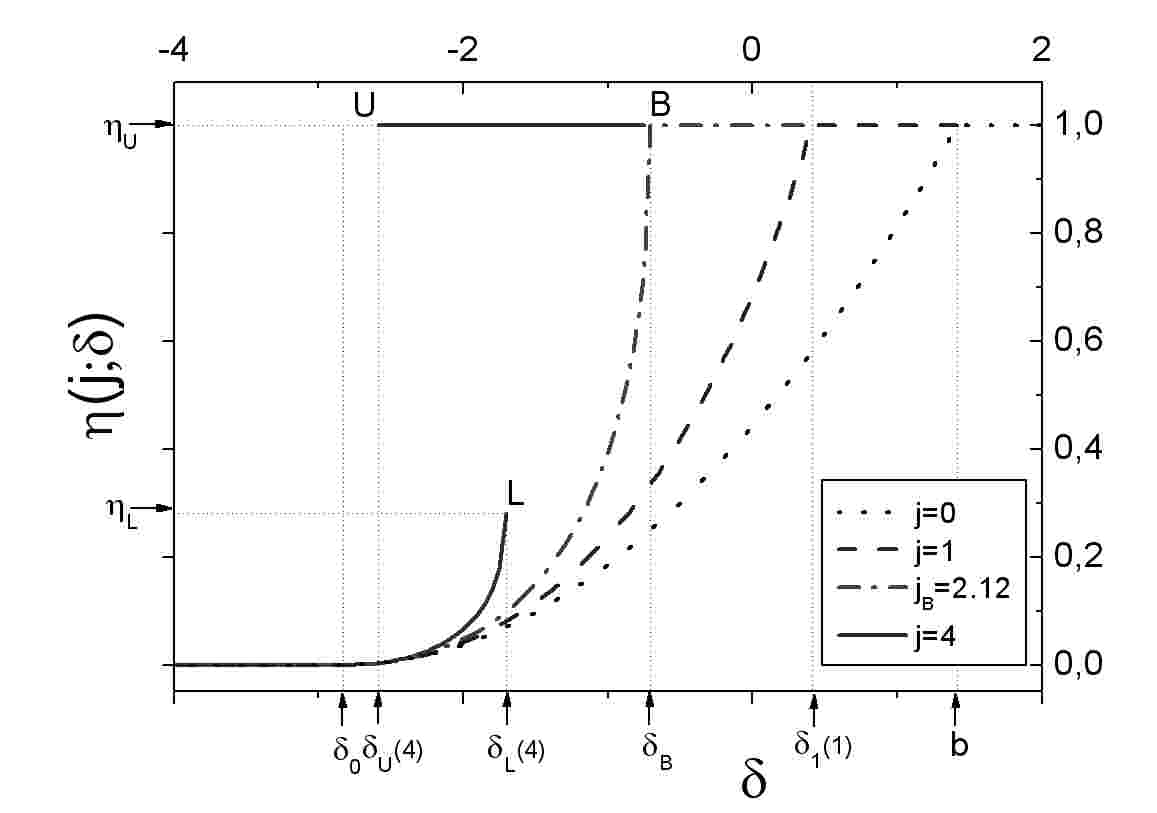} 
\caption[$\eta(\delta)$ as a function of $\delta\equiv h-p$]{ \scriptsize
{Demand $\eta(j;\delta)$ as a function of $\delta\equiv h-p$ for different 
values of $j$, for the triangular pdf 
\protect(\ref{eq:triang_distribution}) with maximum at $x=-b$. Notice that 
$\delta=0$ corresponds to $h=p$. Unstable solutions are not shown. $B$: 
bifurcation point, $U,L$: boundaries of the region with multiple equilibria 
for $j=4$. }}
\label{fig.etaj1j4}
\end{figure}

In our simulations, the reduced variables $x_i$ are randomly 
distributed according to the following triangular pdf:
\begin{equation}
\label{eq:triang_distribution}
f(x)=\frac{2(2b-x)}{9b^2} \;\;\; {\rm if} \;\;\;  -b \leq x \leq 2b, 
\end{equation}
with $b\equiv \sqrt{2}$. Outside the support $[-b,2b]$, $f(x)=0$. 
The maximum of $f(x)$ is reached at the left 
boundary, $f_{max}=f(-b)=2/(3b)$. The solutions to 
(\ref{eq.etaselfconsistent}) are straightforward \cite{PhaSe07},
and are represented as a function of $\delta$ for different values 
of $j$ on figure \ref{fig.etaj1j4}. 

The critical value of $j$ is $j_B=1/f_{max} \approx 2.12$. 
For $j<j_B$, $\eta(j;\delta)$ is a single-valued 
function (see figure \ref{fig.etaj1j4} for the particular 
value $j=1<j_B$). If $\delta <\delta_0 \equiv -2b$ prices 
are so high with respect to the average willingness to pay 
of the population that there are no buyers at all ($\eta=0$), 
i.e. there is no market. At the other end, if 
$\delta > \delta_1(j) \equiv b-j$, prices are so low that the 
market saturates ($\eta=1$). These saturation 
effects arise because the support of $f$ is finite. 
For $\delta_0 \leq\delta\leq \delta_1(j)$, $\eta(j;\delta)$\footnote
{We use the convention that the first terms in parenthesis are parameters, 
and the term after the semicolon is the variable.} is a monotonically 
increasing function of $\delta$:
\begin{equation}
\label{eq:eta_below_B}
\eta(j;\delta)=\frac{9 b^2}{2 j} [1-\sqrt{1-\frac{4j(\delta+2b)}{9b^2} }]-
\frac{(\delta+2b)}{j}.
\end{equation}

For $j>j_B$ there is a range of values of $\delta$, $\delta_U(j) \leq 
\delta \leq \delta_L(j)$ with $\delta_L(j)=-2b + j_B^2/j$ and 
$\delta_U(j)=b-j$ for which there are two solutions\footnote{Notice that 
$\delta_U(j)=\delta_1(j)$ is a degeneracy due to the fact that the pdf 
reaches its maximum at a boundary of the support.}, that we denote 
$\eta_U(j;\delta)$ and $\eta_L(j;\delta)$, with $\eta_U(j;\delta) > 
\eta_L(j;\delta)$ for all the range of $\delta$ where they coexist (see 
$\eta(j;\delta)$ on figure \ref{fig.etaj1j4} for $j=4>j_B$). More 
precisely, the low-$\eta$ branch, $\eta_L(j;\delta)$, exists for 
$\delta<\delta_L(j)$. Its dependence with $j$ and $\delta$ is the same as 
in equation (\ref{eq:eta_below_B}). At $\delta = \delta_L(j)$ it reaches 
its largest value: $\eta_L(j;\delta_L(j))\equiv \eta_L(j)$. The high-
$\eta$ branch exists for $\delta>\delta_U(j)$. In our case, it corresponds 
to saturation ($\eta_U(j;\delta)=1$). The Pareto-optimal equilibrium is 
$\eta_U(j;\delta)$, since it corresponds to the largest utility for all 
the buyers, which are in turn more numerous than in the equilibrium 
$\eta_L(j;\delta)$. However, the equilibrium actually reached by the 
system depends on the decision making process, which we study in the next 
section.

\begin{figure}[ht!]
\centering     
\includegraphics[width=10cm]{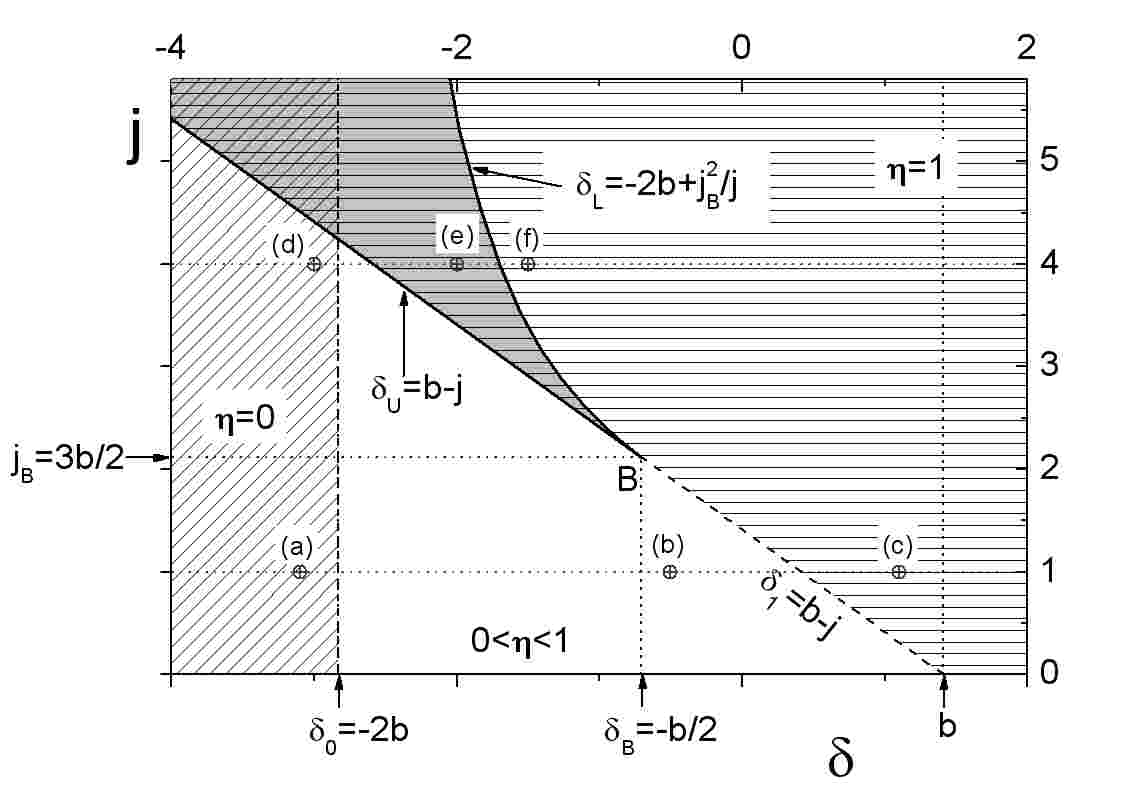} 
\caption[Customers phase diagram.]{ \scriptsize{Customers phase diagram 
for the triangular pdf \protect(\ref{eq:triang_distribution}). Grey 
region: coexistence of two equilibria. $\eta=0$ is an equilibrium within 
the obliquely-hashed region ($\delta <\delta_0$), $\eta=1$ (saturation) is 
an equilibrium within the horizontally-hashed region. Points $(a)$ to $(f)$ 
refer to the parameters considered in section \ref{sec:simulations}.}}
\label{fig.phasediagram}
\end{figure}

These results are summarized on the phase diagram of figure 
\ref{fig.phasediagram} where the saturation lines and the parameter region 
with two solutions (grey area) are represented. 

\section{Learning dynamics}
\label{sec:learning dynamics}
We are interested in the equilibria reached by the system when the 
customers make their decisions repeatedly, at successive periods, based on 
information grasped from their past actions. We assume that at each period 
the agents do not know {\em a priori} the payoffs corresponding to each 
possible strategy. They rely on their own beliefs or estimations to make 
their decisions.  

In the present case of binary decisions it is sufficient to estimate the 
difference between the payoffs expected upon buying with respect to 
not-buying. Thus, individuals need to estimate a single value, that we 
call hereafter attraction for buying, or simply {\em attraction} following 
Camerer \cite{Camerer03}. 

We consider two different learning scenarios which differ in the kind of 
information assumed to be available to the customers. In the first one -- 
that we call $s$-learning --, customers do not know the parameters nor the 
structure of the surplus function on which they have to make expectations. 
They make direct estimations of the payoffs expected upon buying (in our 
model the expected payoff for not buying vanishes). Starting with some 
initial beliefs $a_i(0)$, at each iteration $t$ individuals $i \in \{1, 
\dots, N\}$ make their decisions $\omega_i(t)$ for the period according to 
the attractions $a_i(t)$ and then update the latter based on the obtained 
payoffs $s_i(t)$. In the second scenario -- called hereafter $\eta$
-learning --, each agent is assumed to know the gap between his 
idiosyncratic willingness to pay and the price ($\delta+x_i$) as well as 
the strength of the social interactions $j$. He only needs to estimate the 
expected fraction of buyers $\hat\eta_i(t)$, in order to determine his 
attraction for buying $a_i(t)=\delta+x_i + j \,\hat\eta_i(t)$. 

The system is updated iteratively: at each period $t$ each agent $i$ 
chooses a strategy $\omega_i(t)$ based on his attraction $a_i(t)$. This 
choice may be probabilistic, but here we concentrate on a deterministic 
decision making process. Once decisions are made, attractions are updated 
using the grasped information. More precisely, the system evolves 
according to the following two-steps dynamics:

\paragraph{Making decisions:} each individual makes the choice 
that maximizes his expected payoff. Thus, if the attraction 
for buying is positive, the choice is $\omega_i(t)=1$, 
otherwise $\omega_i(t)=0$. This is called myopic best response 
in the literature. Since attractions are estimated 
payoffs, 
\begin{equation}
\label{eq:best_response}
\omega_i(t)=\Theta (a_i(t)). 
\end{equation}
where $\Theta(x)$ is the Heaviside function 
($\Theta(x)=1$ if $x \geq 0$, $\Theta(x)=0$ otherwise). Notice that this 
deterministic decision rule depends only on the sign of the attraction but 
not on its magnitude. The surplus or earned payoff is then 
\begin{equation}
\label{eq.actual_payoff}
s_i(t) = \omega_i(t) (\delta + x_i +j \eta(t))
\end{equation} 
where $\eta(t)$ is the actual fraction of buyers 
of the period. Since attractions 
may be inaccurate or erroneous estimations of the latter, the agents may 
make bad decisions and either get negative payoffs or miss positive ones.  

\paragraph{Updating attractions:} be $z_i$ the quantity on which the individuals make estimations ($s_i$ or 
$\eta_i$, depending on the learning scenario). Individual $i$ updates 
$\hat z_i(t)$, the estimation at time $t$, using the information obtained 
as a result of his decision $\omega_i(t)$. The updating rules considered 
hereafter have the following structure:
\begin{equation}
\label{eq.WBL}
\hat z_i(t+1) = (1-\mu)\, \hat z_i(t)+
                \mu[\Delta+(1-\Delta)\omega_i(t)]z_i(t)
\end{equation} 
where $0<\mu<1$ is the learning rate and $\Delta$ is a parameter 
($0\leq\Delta\leq1$) that allows to update differently $\hat z_i$ 
depending on the period's decision $\omega_i(t)$. Notice that $z_i(t)$ in 
the right hand side of (\ref{eq.WBL}) is the actual value of $z(t)$ {\em 
after} the decision $\omega_i(t)$ of period $t$ is made and the 
corresponding payoff (if any) is earned. In particular, the learning rule 
obtained by setting $\Delta=1$ in (\ref{eq.WBL}) is known in the literature 
as \emph{fictitious play} \cite{Camerer03, CheungFriedman97}: 
unconditionally to $\omega_i(t)$, the value $\hat z_i(t+1)$ is updated 
using the actual value $z_i(t)$. If $\Delta=0$, the rule (\ref{eq.WBL}) 
gives raise to the usual \emph{reinforcement learning} \cite{ErevRoth98, 
SarinVahid01}, in which the estimated quantity $\hat z$ is updated only if 
$\omega_i(t)=1$. Another well known rule, the standard \emph{Cournot best 
reply} \cite{Cournot60}, in which only the previous period counts, is 
obtained putting $\mu=1$ and $\Delta=1$ in (\ref{eq.WBL}). The latter 
corresponds to a standard parallel steepest ascent search of the 
(eventually local) optimum.

In the $s$-learning scenario, introducing $z_i(t) = s_i(t)$ with $s_i(t)$ 
given by equation (\ref{eq.actual_payoff}), and $\hat z_i 
= a_i$ in equation (\ref{eq.WBL}) gives the time evolution of the 
attraction: 
\begin{equation}
\label{eq.s_learning}
a_i(t+1) = (1-\mu)\, a_i(t)+
                \mu \, [\Delta+(1-\Delta)\omega_i(t)] \, s_i(t)
\end{equation}

In the case of $\eta$-learning, $z_i = \eta_i$ and 
$\hat z_i = \hat \eta_i$, so that, after introduction into 
equation (\ref{eq.WBL}) and some algebra, the evolution 
of the corresponding attraction is:
\begin{equation}
\label{eq.eta_learning}
	a_i(t+1)=(1-\mu)\,a_i(t)+\mu \, (\delta + x_i + j\,[\Delta+(1-\Delta)\omega_i(t)] \, \eta(t))
\end{equation}
Both rules coincide within the fictitious play paradigm, i.e. for 
$\Delta=1$.

\section{General simulation-settings}
\label{sec:setting}
In this section we present the common general settings of 
our simulations. Results obtained with the two different 
learning scenarios presented in the preceding section, 
namely $s$-learning and $\eta$-learning, are discussed in 
the next section. 

\subsection{Systems parameters}
\label{sec.Systems parameters}
Simulations were done for different values of $\delta$, defined by 
(\ref{eq.normalized}).
The values of $x_i$, the (centered) idiosyncratic component of the 
willingness to pay, are drawn according to the triangular pdf (\ref
{eq:triang_distribution}). Since this pdf is a decreasing function of  
$x_i$, there are fewer individuals with high than with low values of 
$x_i$. As a consequence, our histograms of final states as a function of 
$x_i$ have better statistics for low values than for large values of $x_i$. 

We focus on the learning behavior for two values of the social influence 
weight $j$, one below, the other above, the critical value $j_B=3b/2 
\approx 2.12$ (see section \ref{sec:model}). These are $j=1$ which has a 
single equilibrium for any value of $\delta$, and $j=4$, which may present 
two possible equilibria for the range $\delta_U(j) < \delta < \delta_L(j)$ 
with $\delta_U(4)=-2.5858$ and $\delta_L(4)=-1.7034$. At equilibrium, due 
to the boundedness of the support of the IWP, $\eta=0$ below 
$\delta_0\approx 2.83$. For $j=1$ we have $\eta=1$  above 
$\delta_1(1)\approx 0.41$, whereas for $j=4$ saturation ($\eta=1$) is a 
possible equilibrium for $\delta > \delta_U(4)$.

All the presented simulations correspond to systems with $N=1\,000$ 
agents, averaged over $100$ systems, i.e. corresponding to $100$ different 
realizations of the random idiosyncratic willingnesses to pay (IWP). We 
present results corresponding to synchronous (parallel) updating, where 
the procedure detailed in the preceding section is iterated until 
convergence. Results with sequential asynchronous dynamics 
\cite{SeGoNaPh06} only differ in the time needed to converge, the reached 
equilibria being similar. 

We performed thorough simulations, obtaining statistics of learning times, 
cumulated payoffs, etc. In this article we describe the most interesting 
results, which are the fractions of buyers and the distribution of 
attractions at convergence, because they allow to understand the 
differences between the different types of learning schemes.

\subsection{Initial states}
We assume that the agents start with some initial values of 
their attractions, which represent their \emph{a priori} 
beliefs. Among the different possibilities of defining 
the initial beliefs, we analyze systematically three 
different initializations:
\begin{description}
\item{\bf optimistic:} in $s$-learning, the initial attractions $a_i(0)$ 
are randomly selected positive numbers in the interval $[0,1]$ for all $1 
\leq i \leq N$, so that the very first decision for all the agents is to 
buy. In $\eta$-learning, the initial values are $\hat \eta_i(0)=1$ for all 
$1 \leq i \leq N$, so that the initial attractions are 
$a_i(0)=x_i+\delta+j$.
Notice that in this case, the decisions of agents with $x_i<-j-\delta$ is 
not strategic: they will choose not to buy in the first period despite 
their optimistic guess on $\hat \eta_i(0)$, because their IWP is too small. 

\item{\bf pessimistic:} in $s$-learning, the 
initial attractions $a_i(0)$ are randomly selected negative 
numbers in the interval $[-1,0]$ for all  $1 \leq i \leq N$. 
At the first iteration, no agent buys.
In $\eta$-learning, the initial values are 
$\hat \eta_i(0)=0$ for all  $1 \leq i \leq N$. 
Here, the choices of individuals with $x_i>-\delta$ are not strategic 
because their payoffs upon buying are positive independently of the 
choices of the other agents. Thus, their first period choice is 
$\omega_i(0)=1$.

\item{\bf random:} in $s$-learning, the initial attractions $a_i(0)$ are 
randomly selected numbers in the interval $[-1,1]$ for all $1 \leq i \leq 
N$. In $\eta$-learning, the initial values $\hat \eta_i(0)$ are random 
numbers in $[0,1]$ for all  $1 \leq i \leq N$. Here, agents with 
$x_i+\delta>0$ (resp. $x_i+\delta-j<0$) buy (resp. do not buy) 
unconditionally to the individual estimations $\hat\eta_i(0)$. 
Those with $-j<x_i+\delta<0$, i.e. those whose decision is actually 
dependent on the collective outcome, will buy only if 
$\hat\eta_i(0)>-(\delta+x_i)/j$.
\end{description}
The two first initializations correspond to extreme cases. 
They lead to equilibria that are respectively upper and lower bounds to the 
fractions of buyers at equilibrium reached with other initializations.

\section{Simulations results}
\label{sec:simulations}
We first present results obtained with myopic fictitious play, for which 
both learning scenarios coincide. This corresponds to 
the usual dynamics used in spin systems, in which at each iteration spins 
are aligned with their local fields. Since the interactions between agents 
are symmetric, the system has an underlying energy function. Thus, the 
dynamics has fixed point attractors, which are the equilibrium states 
presented in section \ref{sec:model}. These results will serve as 
reference against which we compare the results of weighted belief and 
reinforcement learning. 

\subsection{Myopic fictitious play}
\label{sec.Myopic fictitious play}
This dynamics is achieved by putting $\Delta=1$ and $\mu=1$ in equation 
(\ref{eq.WBL}). It is called myopic because it is a response to the 
previous time step only: agents completely disregard older experiences and 
do not try to make elaborate expectations on future outcomes. Fictitious 
because agents are assumed to have knowledge of the values ($s$ or $\eta$) 
used to build their attractions independently of whether they buy or not. 
 
\begin{figure}[ht!]
\centering     
\begin{tabular}{cc}
\includegraphics[width=7cm]{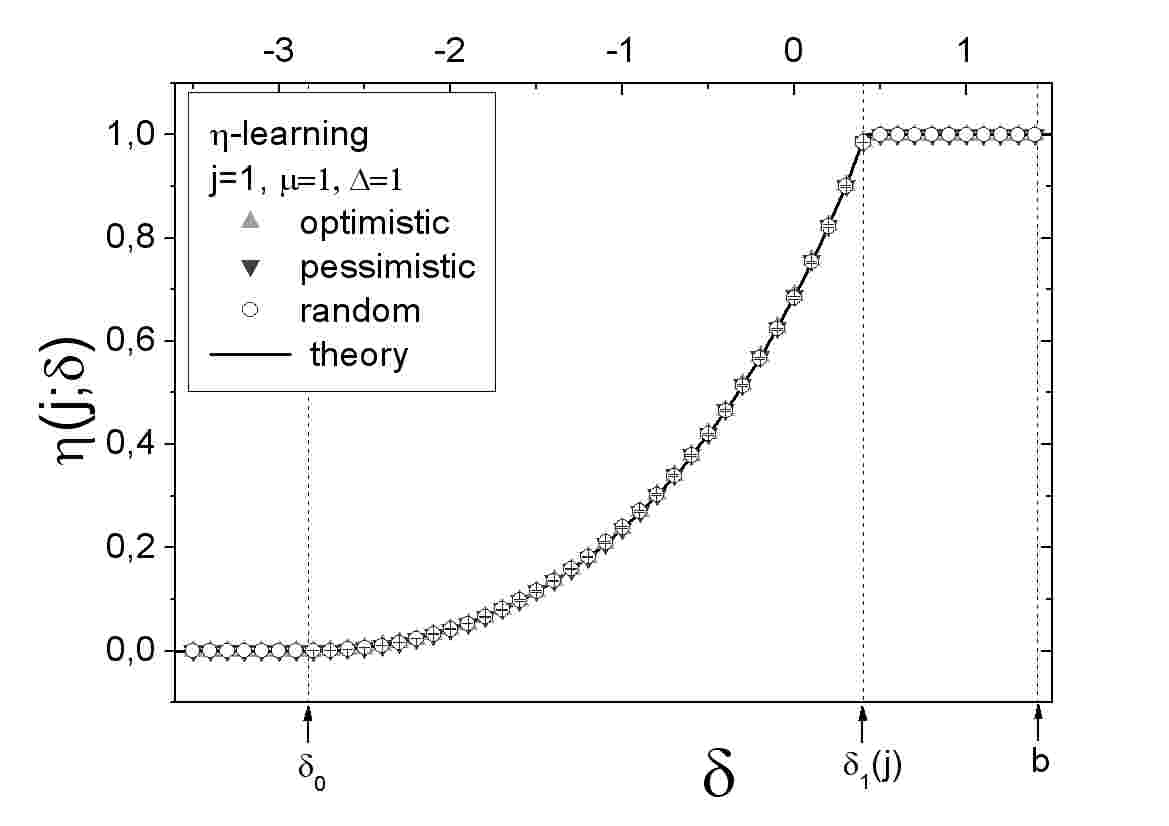} 
\includegraphics[width=7cm]{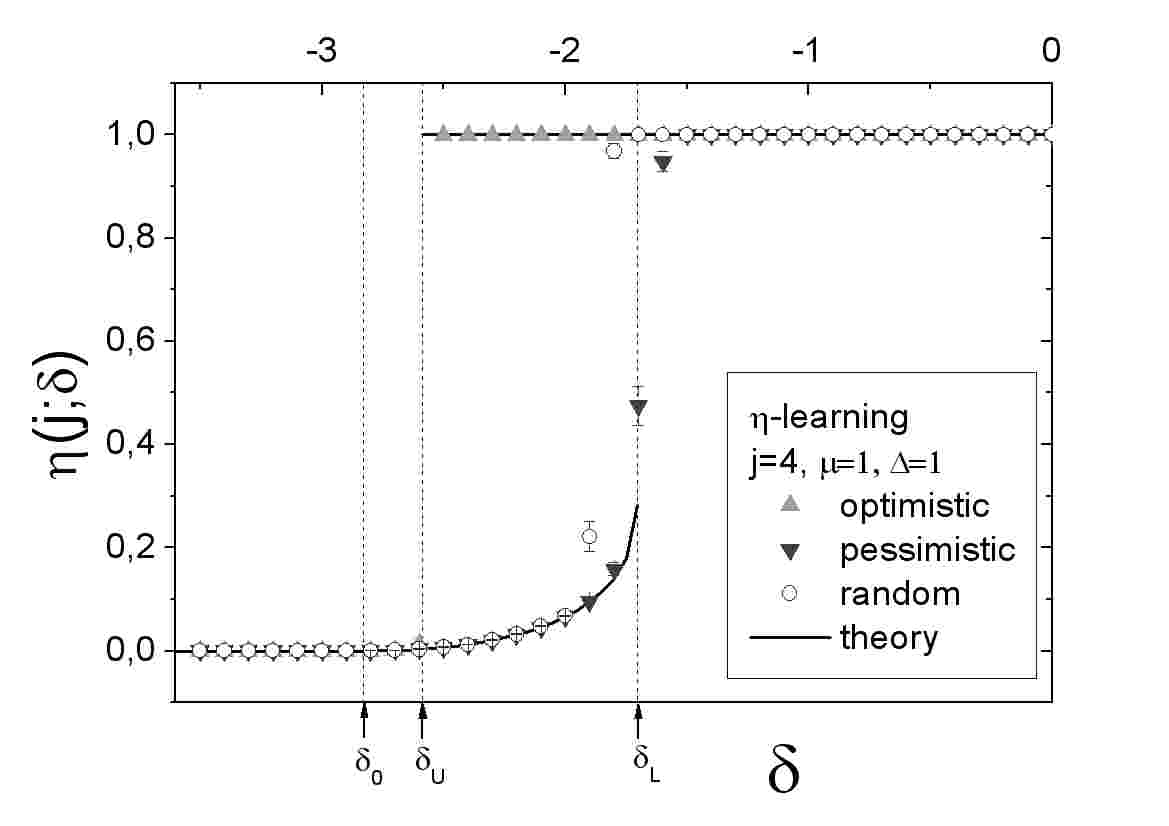} 
\end{tabular}
\caption[Fraction of buyers with Myopic Fictitious Play]{ \scriptsize{Myopic fictitious play ($\mu=1$, $\Delta=1$). $\eta$ at equilibrium versus $\delta\equiv h-p$ for $j=1$ and $j=4$. In these and all the following figures, when non visible, error bars are smaller than the symbols' sizes. The full lines are the analytical predictions for the rational (Nash) equilibria.} Numerical values of $\delta_0$, $\delta_1(j)$, $\delta_U(j)$ and $\delta_L(j)$ are given in section \protect{\ref{sec.Systems parameters}}.}
\label{fig.MBR_eta}
\end{figure}

The fractions of buyers $\eta$ at equilibrium, obtained for different 
values of $\delta$, are presented on figure \ref{fig.MBR_eta}. Symbols 
correspond to simulations, the lines being the solutions to the mean field 
equation (\ref{eq.etaselfconsistent}) with the triangular IWP distribution 
(\ref{eq:triang_distribution}), represented on figure \ref{fig.etaj1j4}. 
In the range $0 < \eta < 1$ (excluding $\eta=0$ and $\eta=1$), these 
solutions are given by equation (\ref{eq:eta_below_B}). 

Figure \ref{fig.MBR_eta} (left) displays results for $j=1<j_B$. This 
dynamics corresponds to steepest ascent in the states space, so that the 
system reaches the fixed point closest to the initial state. Since for 
$j=1$ there is only one fixed point for each value of $\delta$, the system 
converges to it independently of the initialization. For 
$\delta_0\leq\delta\leq\delta_1(j)$, $\eta$ is the fraction 
of agents that satisfy $x_i+\delta+j\eta>0$. In the region 
$\delta>\delta_1(j)$, these are all the agents. If 
$\delta<\delta_0$, no agent has an IWP large enough to get a positive 
payoff, and the equilibrium is $\eta=0$. 

For $j>j_B$ and $\delta_U(j) \leq \delta \leq \delta_L(j)$ we expect, based on 
the phase diagram, that different initializations  lead the system to 
different equilibria. Indeed, the optimistic (pessimistic) initialization 
systematically drives the system to the high-$\eta$ (low-$\eta$) 
equilibrium. Systems with random initialization end up at either of the two 
equilibria, depending on the precise configuration of initial states (see 
figure \ref{fig.MBR_eta}, right). Actually, with this initialization the 
distribution of $\eta$ is bimodal; this is why the averages in the 
coexistence region present larger variances than elsewhere. With different 
initial fractions of buyers, the number of simulated systems that end up at 
each attractor differs. 

\begin{figure}[ht!]
\centering     
\begin{tabular}{cc}
\includegraphics[width=7cm]{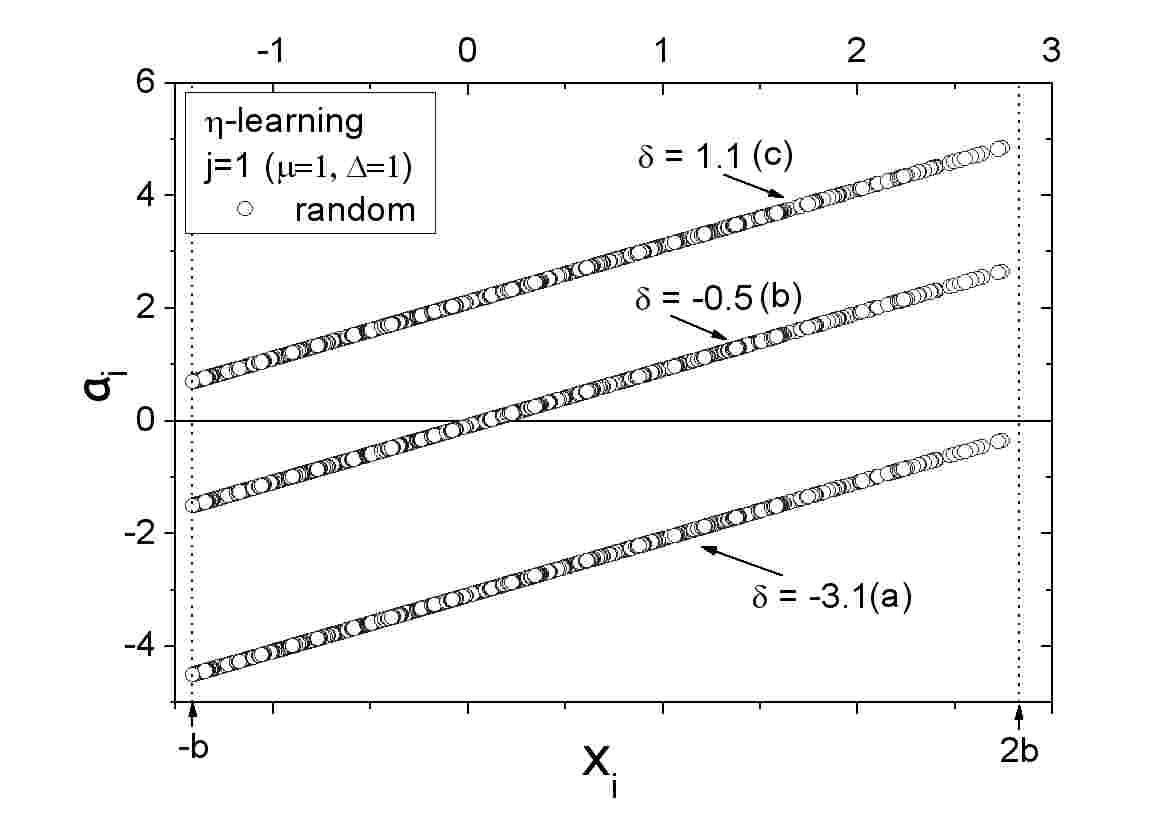}  
\includegraphics[width=7cm]{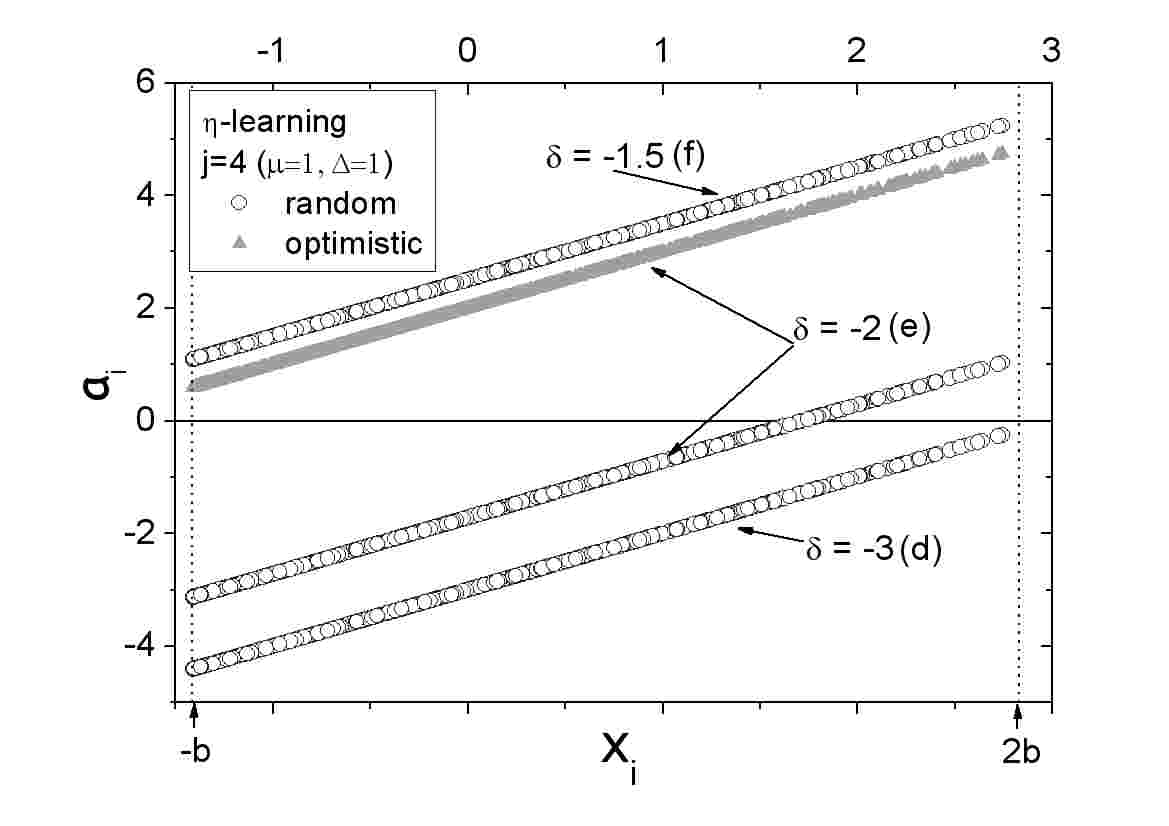} 
\end{tabular}
\caption[]{ \scriptsize{Myopic fictitious play ($\mu=1$, $\Delta=1$). 
Attractions at convergence for three different values of $\delta$, as a 
function of $x_i$. For $j=1$, the fractions of buyers are 
$\eta(\delta=1.1)=1$, $\eta(\delta=-0.5)=0.403$ and $\eta(\delta=-3.1)=0$. 
For $j=4$, $\eta(\delta=-1.5)=1$ and $\eta(\delta=-3)=0$. For $\delta=-2$ 
attractions converge to two different fixed points ($\eta=0.07$ and 
$\eta=1$), depending on the initial condition. The characters in 
parenthesis refer to the points in the phase diagram 
(figure \protect{\ref{fig.phasediagram}}).}}
\label{fig.MBR_a_iwp}
\end{figure}
The stationary distribution of the individual attractions $a_i$ of a single 
typical system are plotted on figures \ref{fig.MBR_a_iwp} against the 
idiosyncratic terms $x_i$, for different values of $\delta$ (they 
correspond to the equilibrium states $(a)-(f)$  in the phase diagram 
\ref{fig.phasediagram}). As expected, the $a_i$ are the actual payoffs at 
equilibrium, which are proportional to $\eta$. The slope of $a_i$ vs. 
$x_i$ is 1, as it should, since $a_i=\delta+x_i+j\eta$, the ordinate at the origin being $\delta+j \eta$. 

Results for $j=4$ and $\delta=-2$ (point (e) in the phase diagram, inside 
the coexistence region $\delta_U(4)\leq \delta \leq \delta_L(4)$) show the two 
possible outcomes, obtained through different initializations, 
corresponding to the two possible fixed points.

\subsection{Weighted belief learning}
In the weighted belief learning scenario, the information grasped by 
buyers has a larger weight than that of non-buyers. This scenario aims at 
modelizing situations where buyers have first hand knowledge of the 
quantities they try to estimate (payoffs or fraction of buyers) whereas 
individuals that do not afford the risk of buying have less faithful 
information. In equation (\ref{eq.WBL}) this is achieved whenever 
$0<\Delta < 1$. 
 
\begin{figure} [ht!]
\centering
\begin{tabular}{cc}
\includegraphics[width=7cm]{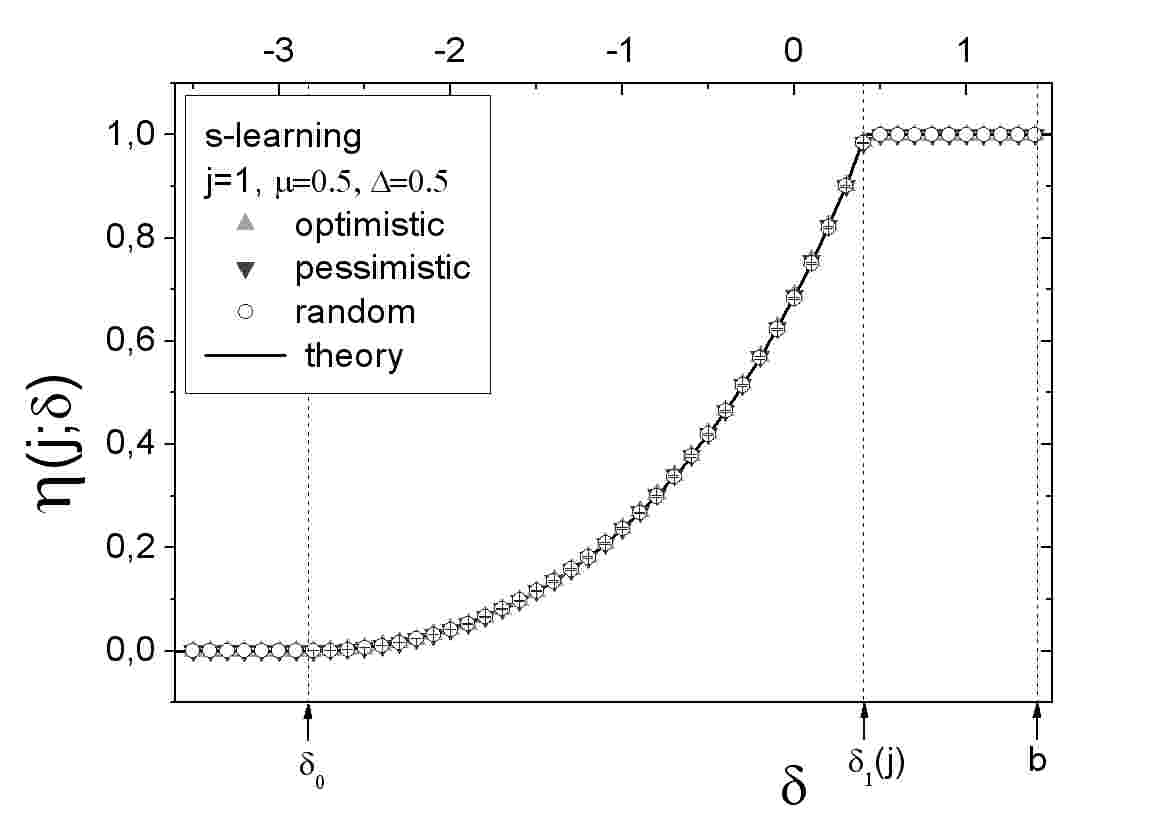} 
\includegraphics[width=7cm]{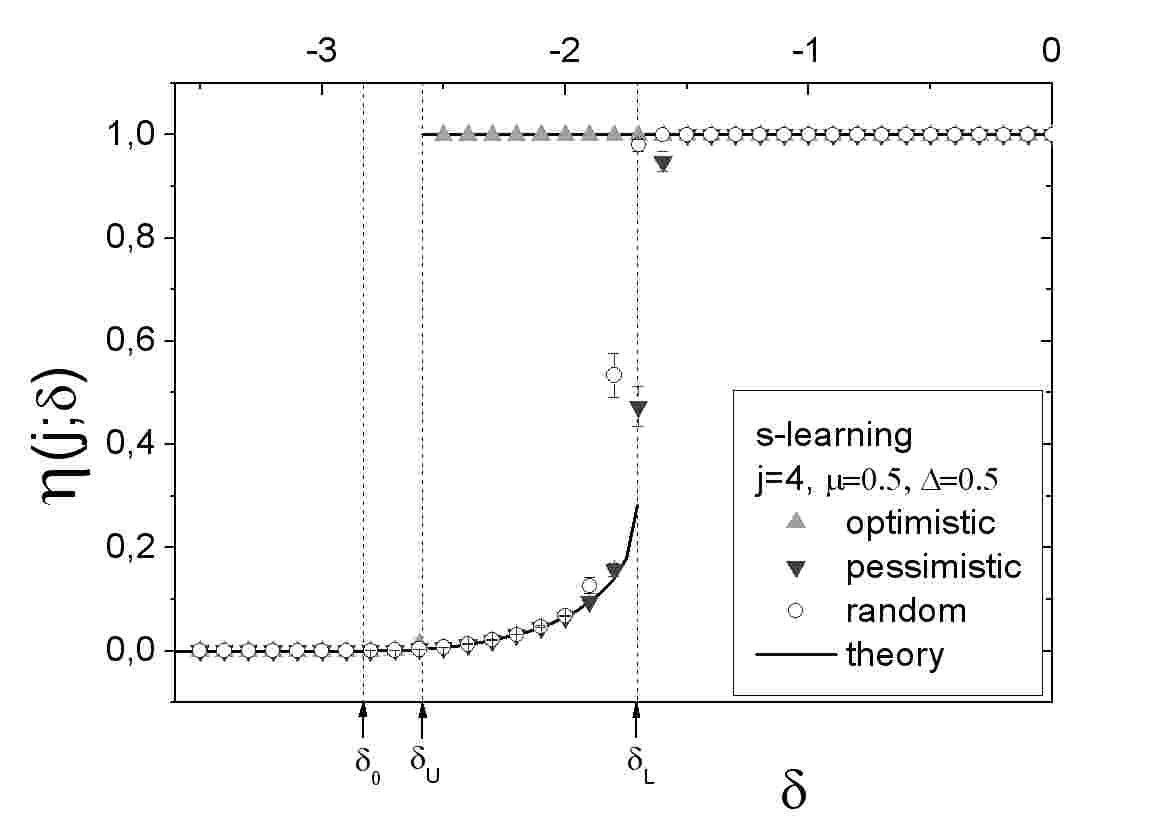} \\
\includegraphics[width=7cm]{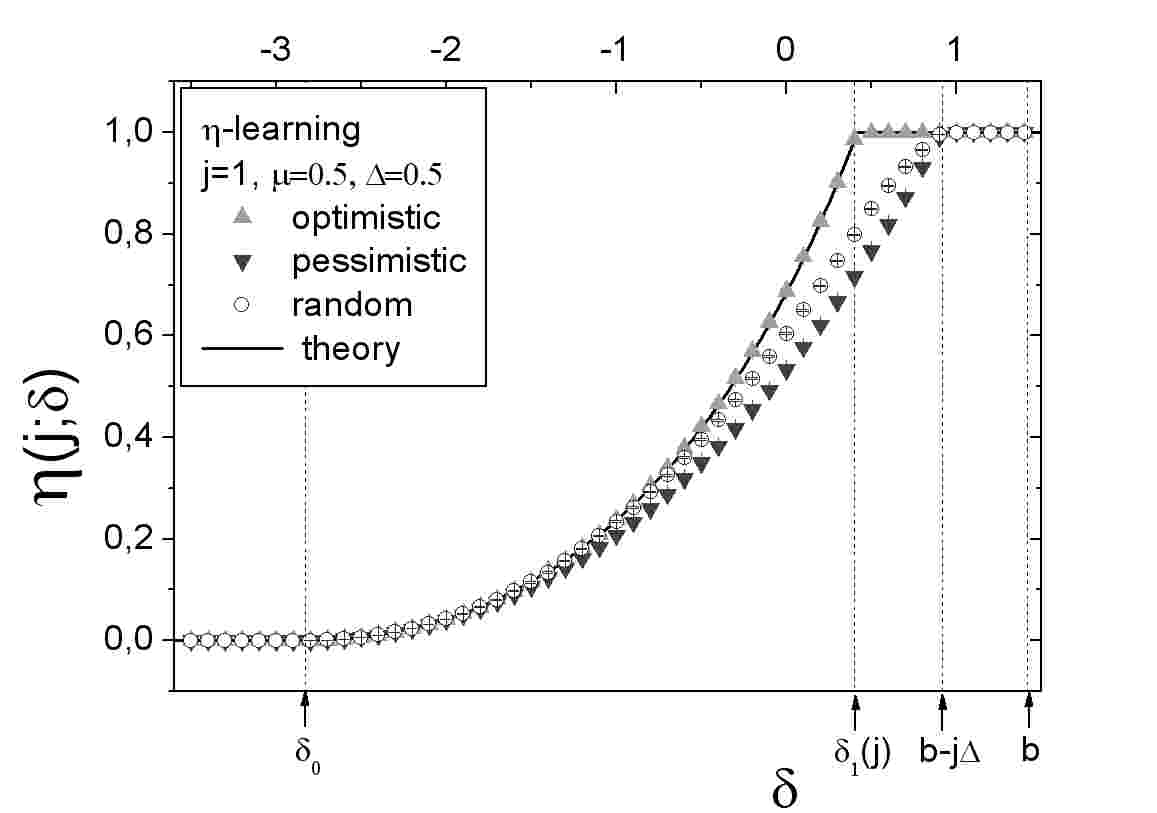}
\includegraphics[width=7cm]{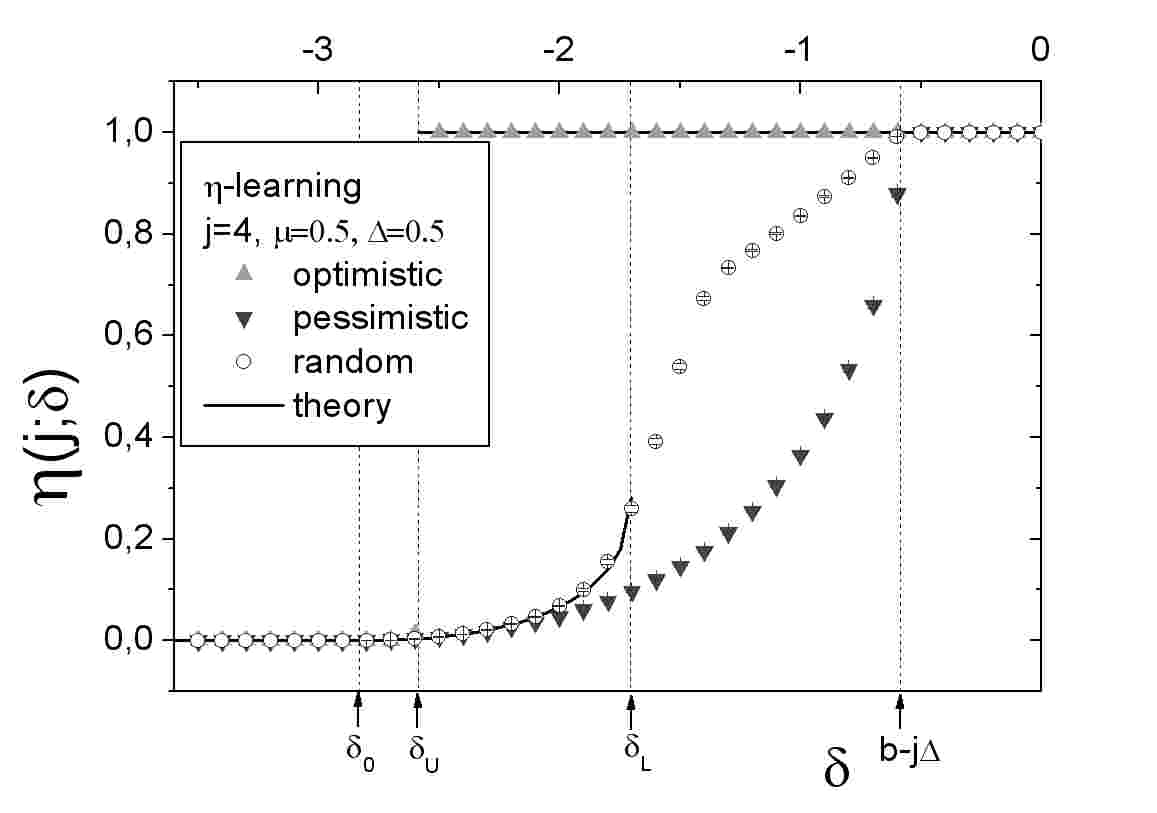}
\end{tabular}
\caption[]{ \scriptsize{Weighted belief learning ($\mu=0.5$, $\Delta=0.5$). $\eta$ at equilibrium versus $\delta$ for $j=1$ and $j=4$, for $s$-learning (above) and $\eta$-learning (below). }}
\label{fig.WBL_aver_eta}
\end{figure}
The fractions of buyers $\eta$ at convergence with $\Delta=0.5$ are 
plotted on figures \ref{fig.WBL_aver_eta}. With $s$-learning, both for 
$j<j_B$ and $j>j_B$, equilibria are similar to those with myopic 
fictitious play, independently of the type of initialization, although we 
show later that the learned attractions are quite different. In contrast, 
the states reached with $\eta$-learning crucially depend on $\Delta$ being 
smaller than $1$. In fact, only with the optimistic initialization the 
agents may reach coordination on the optimal equilibrium (if it exists). 
With the other two initializations non-buyers systematically underevaluate 
the social effects by a factor $\Delta$. As a result, the $a_i$'s are 
underevaluated and the collective outcomes at equilibrium are not 
consistent with the phase diagram. At convergence $\eta$ is smaller than 
the optimal value for a large range of $\delta$ values (see figures \ref
{fig.WBL_aver_eta}). This decrease in $\eta$ is most dramatic with the 
pessimistic initialization. Since with the random initialization there are 
more buyers than with the pessimistic initialization from the beginning, 
more individuals can correctly estimate their surpluses, and the 
collective state at equilibrium has systematically a larger $\eta$ than 
when starting with the pessimistic initialization.

\begin{figure}[ht!] 
\centering
\begin{tabular}{cc}
\includegraphics[width=7cm]{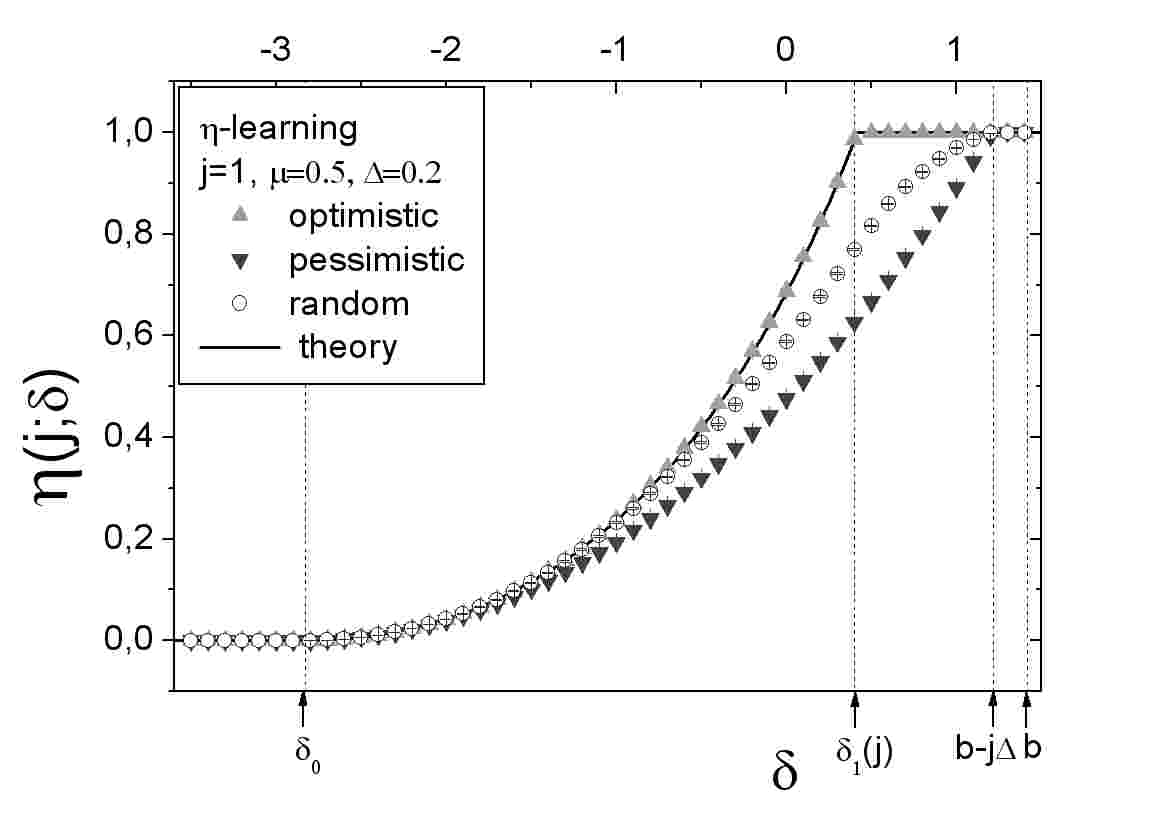} 
\includegraphics[width=7cm]{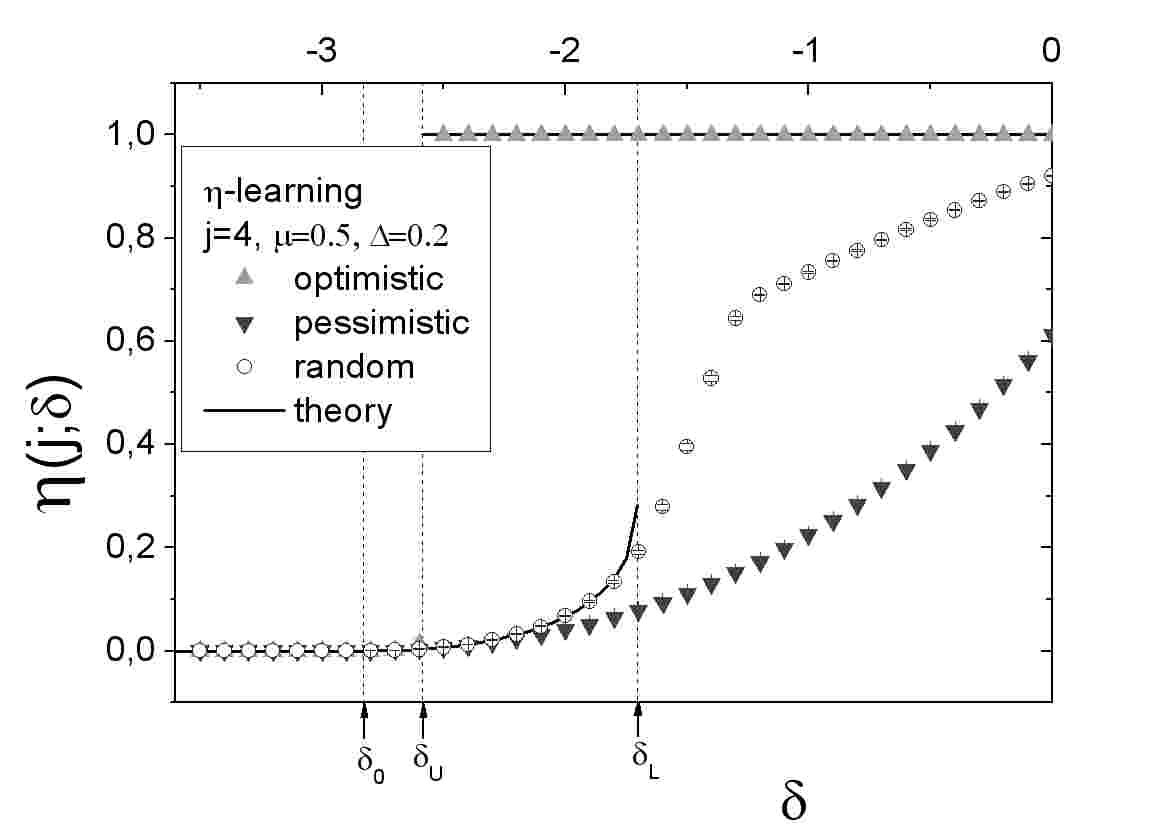}\\ 
\includegraphics[width=7cm]{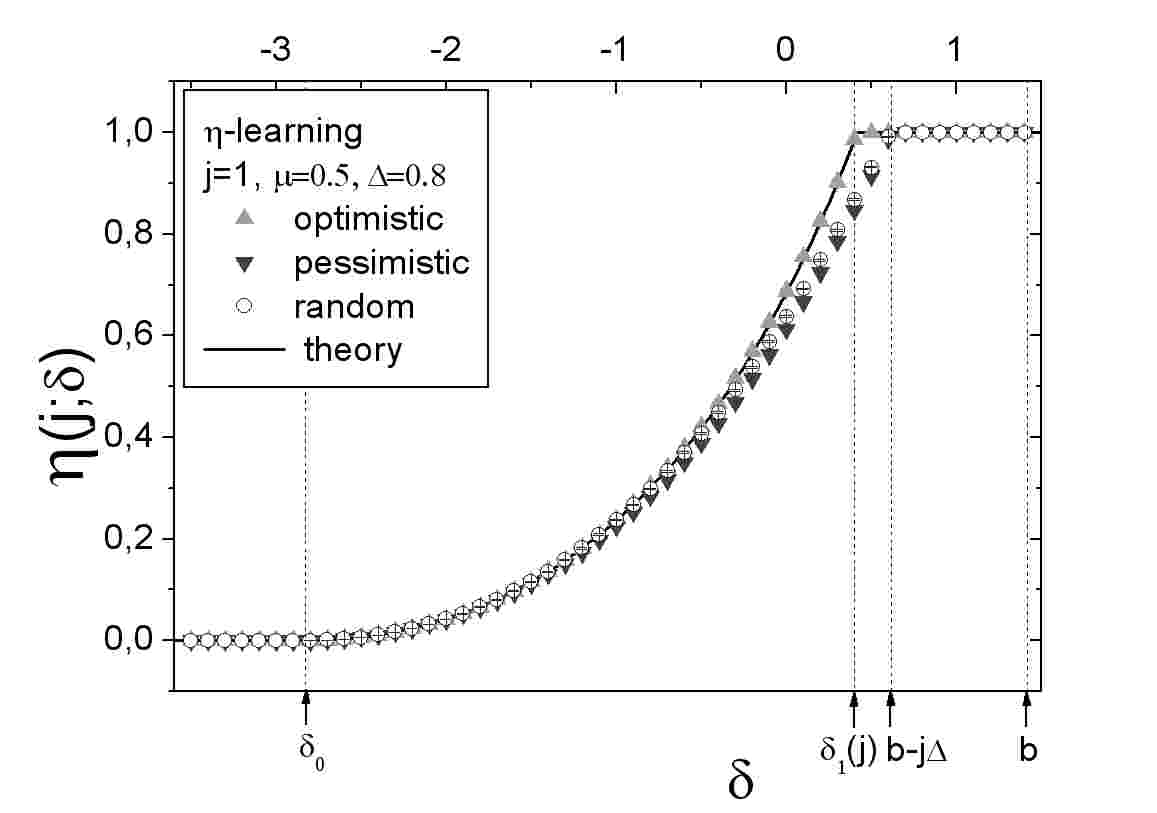} 
\includegraphics[width=7cm]{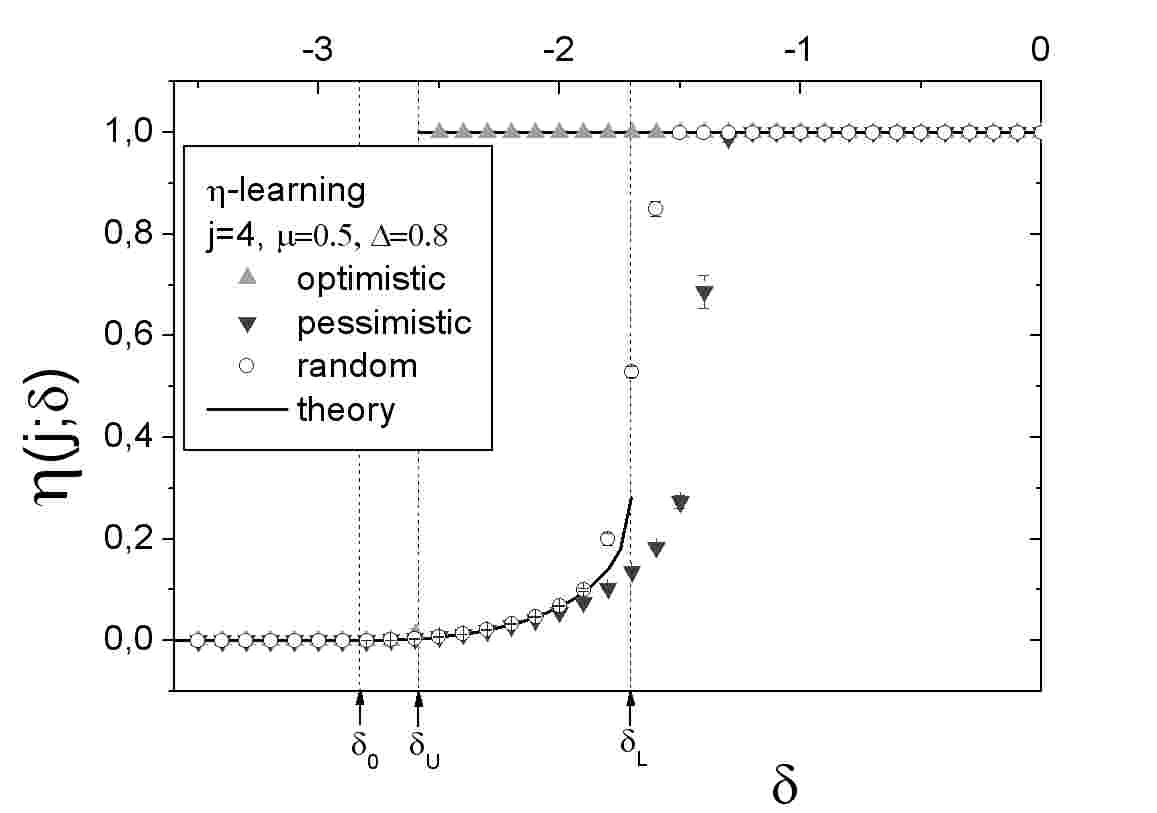}
\end{tabular}
\caption[]{ \scriptsize{Weighted belief learning ($\mu=0.5$): dependence of $\eta$-learning on $\Delta$: $\eta$ at equilibrium versus $\delta$ for $j=1$ and $j=4$ with $\Delta=0.2$ (above) and $\Delta=0.8$ (below).}}
\label{fig.WBL_aver_eta_d}
\end{figure}

For smaller values of $\Delta$, the misestimations of the social terms are 
even more conspicuous, leading the population to inefficient equilibria 
with low fractions of buyers for a larger range of $\delta$ values (see 
figures \ref{fig.WBL_aver_eta_d}, $\Delta=0.2$). Conversely, when $\Delta$ 
is larger, the actual and the estimated $\eta$ are closer to each other, 
giving results closer to those of fictitious play (see figures \ref
{fig.WBL_aver_eta_d}, $\Delta=0.8$). In the limit $\Delta \to 1$ we obtain 
the results of section \ref{sec.Myopic fictitious play}. The case $\Delta=0$
is considered in the next section.

\begin{figure}[ht!]
\centering     
\begin{tabular}{cc}
\includegraphics[width=7cm]{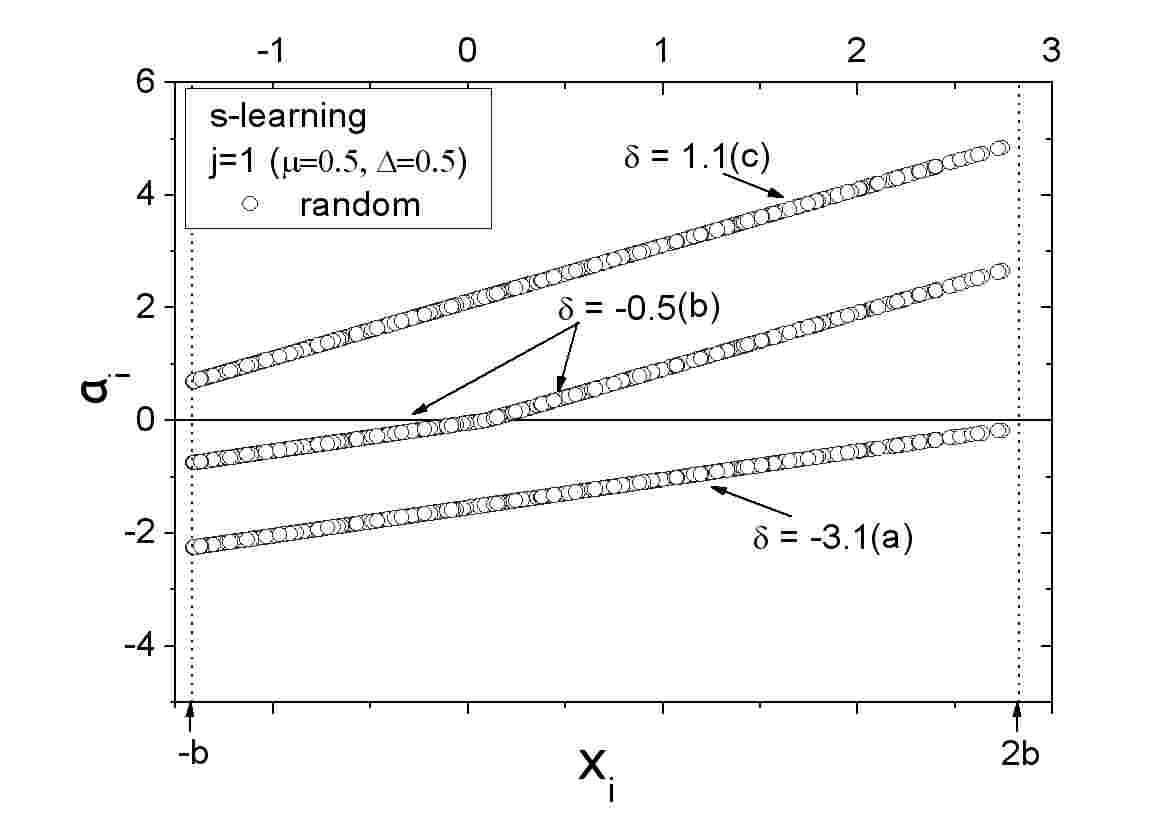} 
\includegraphics[width=7cm]{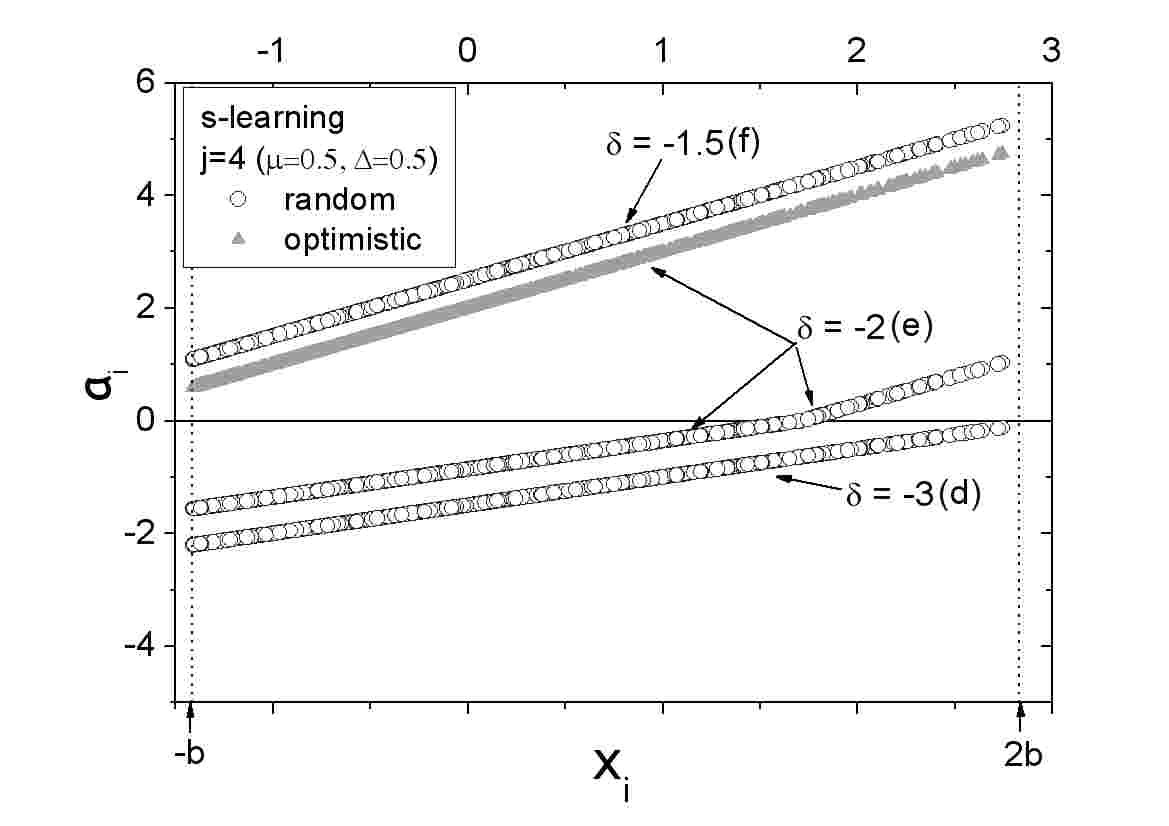} \\
\includegraphics[width=7cm]{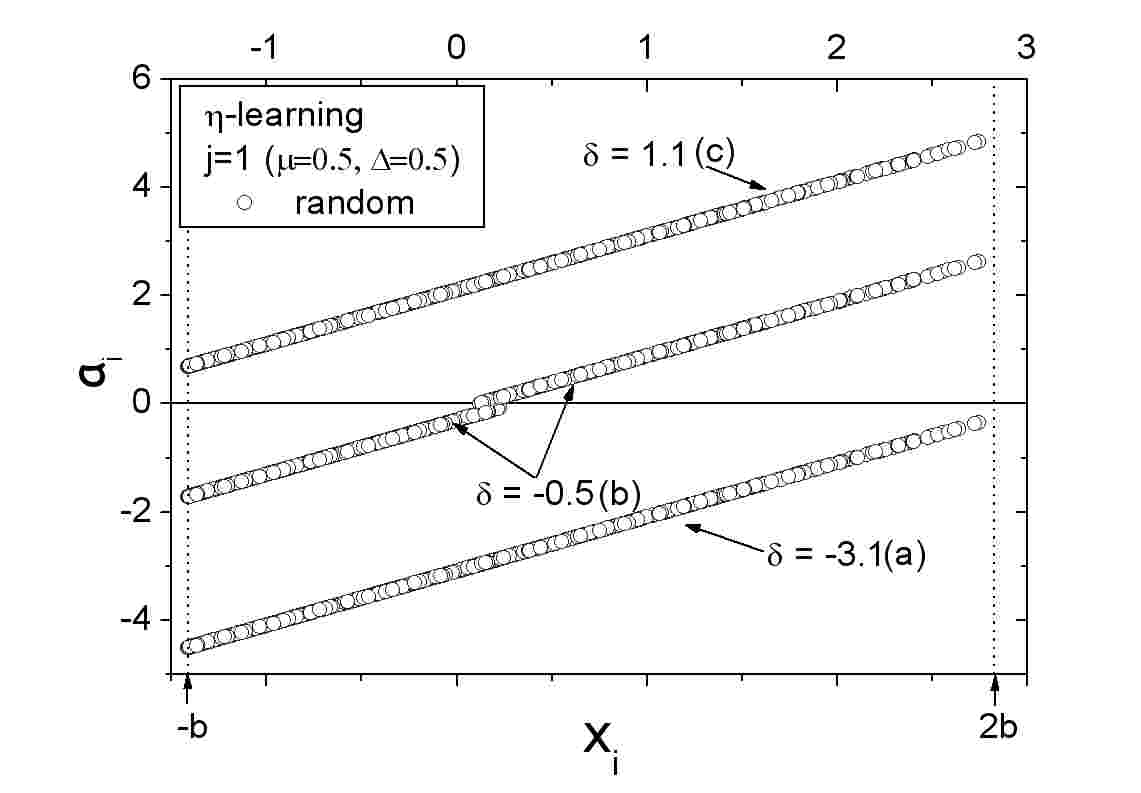}  
\includegraphics[width=7cm]{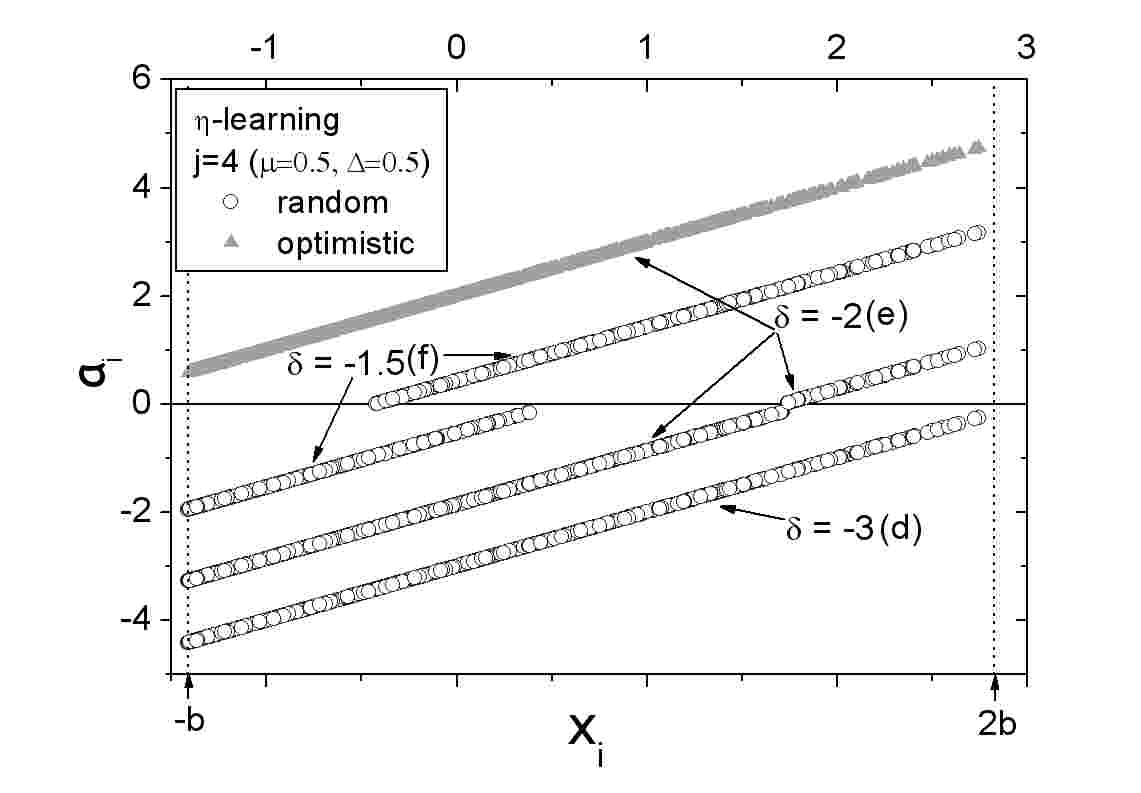} 
\end{tabular}
\caption[]{ \scriptsize{Weighted belief learning ($\mu=0.5$, 
$\Delta=0.5$). Attractions at convergence for three 
values $\delta$, as a function of $x_i$. Above, for $s$-learning: 
$\eta(j=1,\delta=1.1)=1$, $\eta(j=1,\delta=-0.5)=0.403$ and 
$\eta(j=1,\delta=-3.1)=0$. For $j=4$, when $\delta=-2$ attractions 
converge to two different fixed points (with $\eta=0.078$ 
and $\eta=1$), depending on the initial condition) whereas $\eta=1$ for 
$\delta=-1.5$ and $\eta=0$ for $\delta=-3$. Below, for $\eta$-learning: 
$\eta(j=1,\delta=1.1)=1$, $\eta(j=1,\delta=-0.5)=0.382$ and 
$\eta(j=1,\delta=-3.1)=0$. For $j=4$, when 
$\delta=-2$ attractions converge to two different fixed points (with 
$\eta=0.03$ and $\eta=1$, depending on the 
initial condition), whereas $\eta(j=4,\delta=-1.5)=0.62$ and 
$\eta(j=4,\delta=-3=0$ .}}
\label{fig.WBL_a_iwp}
\end{figure}

The individual attractions at convergence of a representative system are 
represented against the individual idiosyncratic terms $x_i$ on figures 
\ref{fig.WBL_a_iwp}, for $j=1$ and $j=4$ and for different values of 
$\delta$. In contrast with myopic fictitious play, the slope of the 
attractions obtained with $s$-learning depends on whether individuals are 
buyers or not: for non-buyers the slope is $\Delta$ whereas it is $1$ for 
buyers, as may be seen on the upper figures \ref{fig.WBL_a_iwp}.

In the case of $\eta$-learning it is clear from the updating rule (\ref
{eq.eta_learning}) that the attractions as a function of $x_i$ have a slope 
$1$. However, because $\Delta < 1$, both with the pessimistic and the random 
initializations the fractions of non-buyers when $j=1$ for $\delta$ in 
the region $\delta_1(j)<\delta<b-j\Delta$ (see figure \ref
{fig.WBL_aver_eta}) do not reach the saturation level expected from the 
phase diagram. The non-buyers are agents whose initial estimations 
$\hat\eta_i(0)$ determined negative attractions. When the correcting term 
$j\Delta\eta$ does not allow  to compensate a negative value of 
$\delta+x_i$, these agents persist in non buying. 

When $j=4$, for $\delta>\delta_L(j)$ there is a fraction of the population 
that does not buy, due to same reason as for $j<j_B$. This is why for 
$\delta=-1.5>\delta_L(4)$, where saturation is expected on the basis of the 
phase diagram, we obtain $\eta<1$ with either random or pessimistic 
initializations. Like for $j=1$, here also saturation is reached 
independently of the initial state only for $\delta > b-j \Delta$.

\subsection{Reinforcement learning}

\begin{figure}[ht!]
\centering     
\begin{tabular}{cc}
\includegraphics[width=7cm]{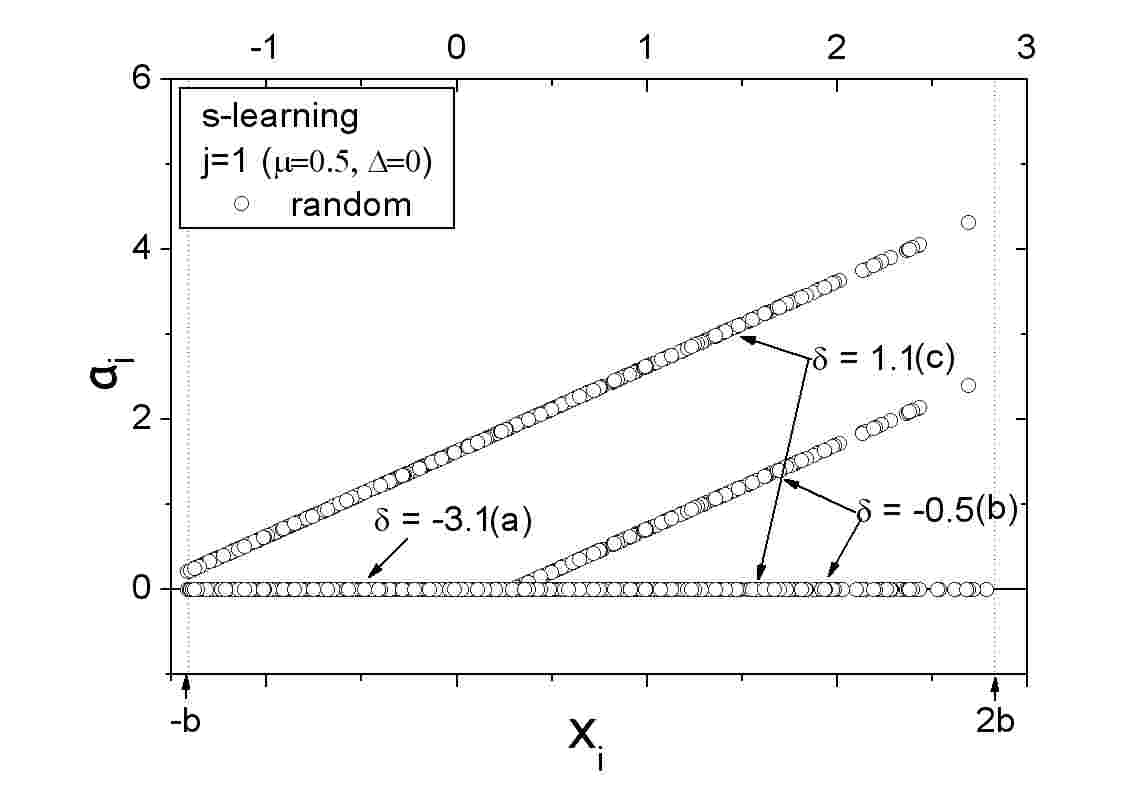} 
\includegraphics[width=7cm]{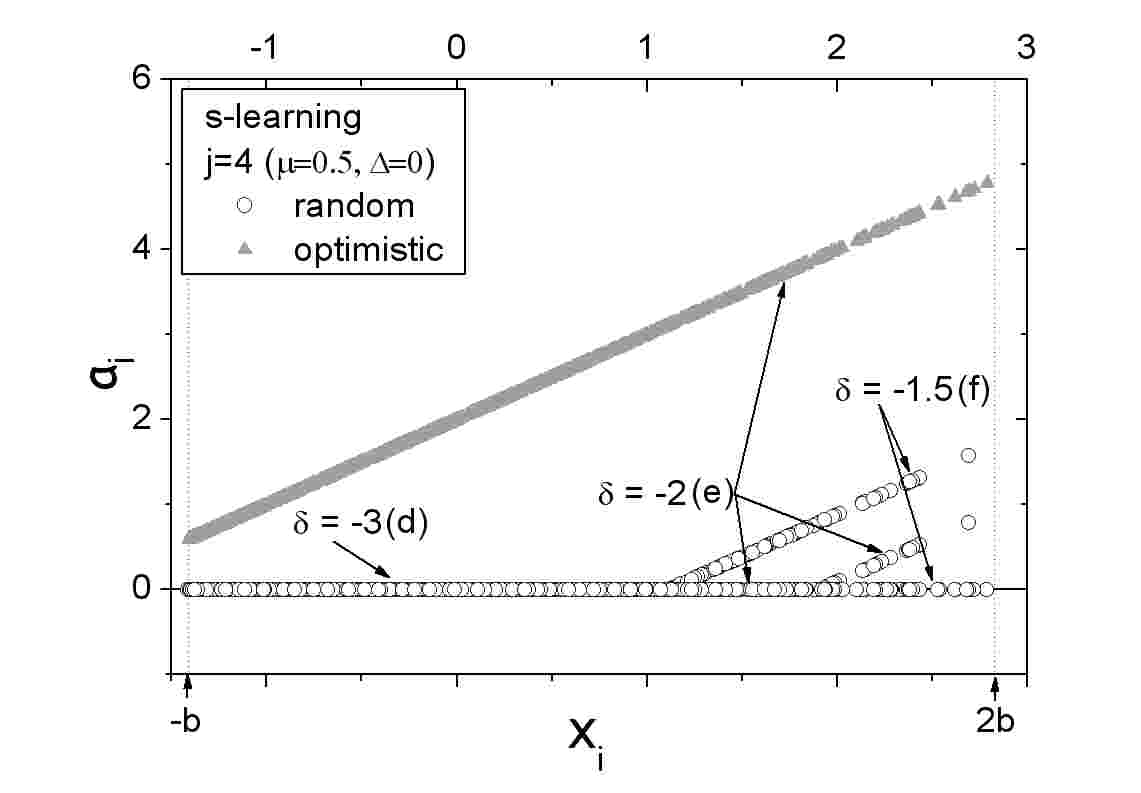}\\
\includegraphics[width=7cm]{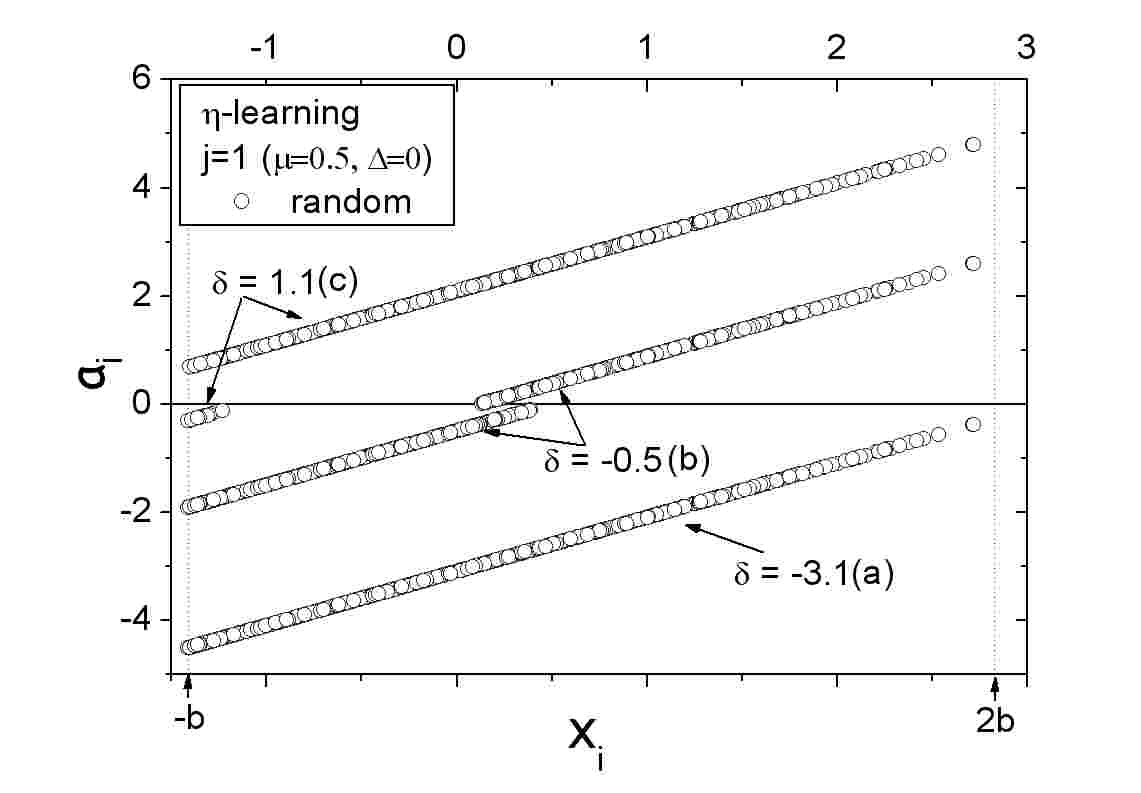} 
\includegraphics[width=7cm]{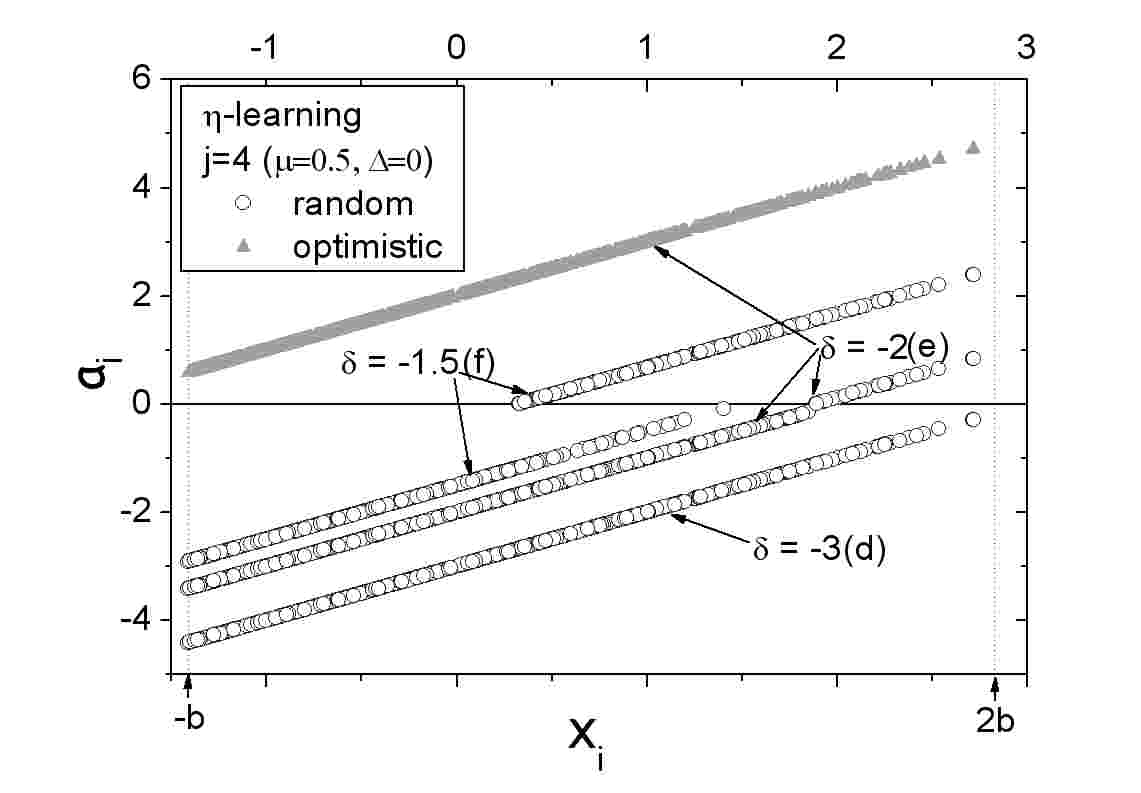} 
\end{tabular}
\caption[]{ \scriptsize{Reinforcement learning ($\mu=0.5$, $\Delta=0$). 
Attractions at convergence for three values of $\delta$, as a function of 
$x_i$. $s$-learning (above): the fractions of buyers for $j=1$ are 
$\eta(j=1,\delta=1.1)=0.52$, $\eta(j=1,\delta=-0.5)=0.203$ and 
$\eta(j=1,\delta=-3.1)=0$. For $j=4$, $\delta=-2$, attractions converge to 
two different fixed points (with $\eta=0.022$ and $\eta=1$, depending on 
the initial condition), whereas $\eta(j=4,\delta=-1.5)=0.095$ and 
$\eta(j=4,\delta=-3)=0$. $\eta$-learning (below): 
$\eta(j=1,\delta=1.1)=0.985$, $\eta(j=1,\delta=-0.5)=0.374$ and 
$\eta(j=1,\delta-3.1)=0$. For $j=4$, $\delta=-2$ attractions converge to 
two different fixed points ($\eta=0.03$ and $\eta=1$, depending on the 
initial condition), whereas $\eta(j=4,\delta=-1.5)=0.294$ and 
$\eta(j=4,\delta=-3)=0$.}}
\label{fig.RL_a_iwp}
\end{figure}

In reinforcement learning, only agents that buy are assumed to have the 
information necessary to estimate their attractions. In equation (\ref
{eq.WBL}) this is achieved with $\Delta=0$. This learning paradigm is 
also called stimulus-response or rote learning in behavioral psychology \cite{BushMosteller55}. 
It aims at modelizing risk-averse individuals that refrain from buying from 
the start, independently of the posted price. As we see in the following, 
such behaviours may hinder the emergence of the Pareto-optimal equilibrium, 
where the payoffs are optimal for all the agents, for a large range of 
values of $\delta$.

\begin{figure} [ht!]
\centering 
\begin{tabular}{cc} 
\includegraphics[width=7cm]{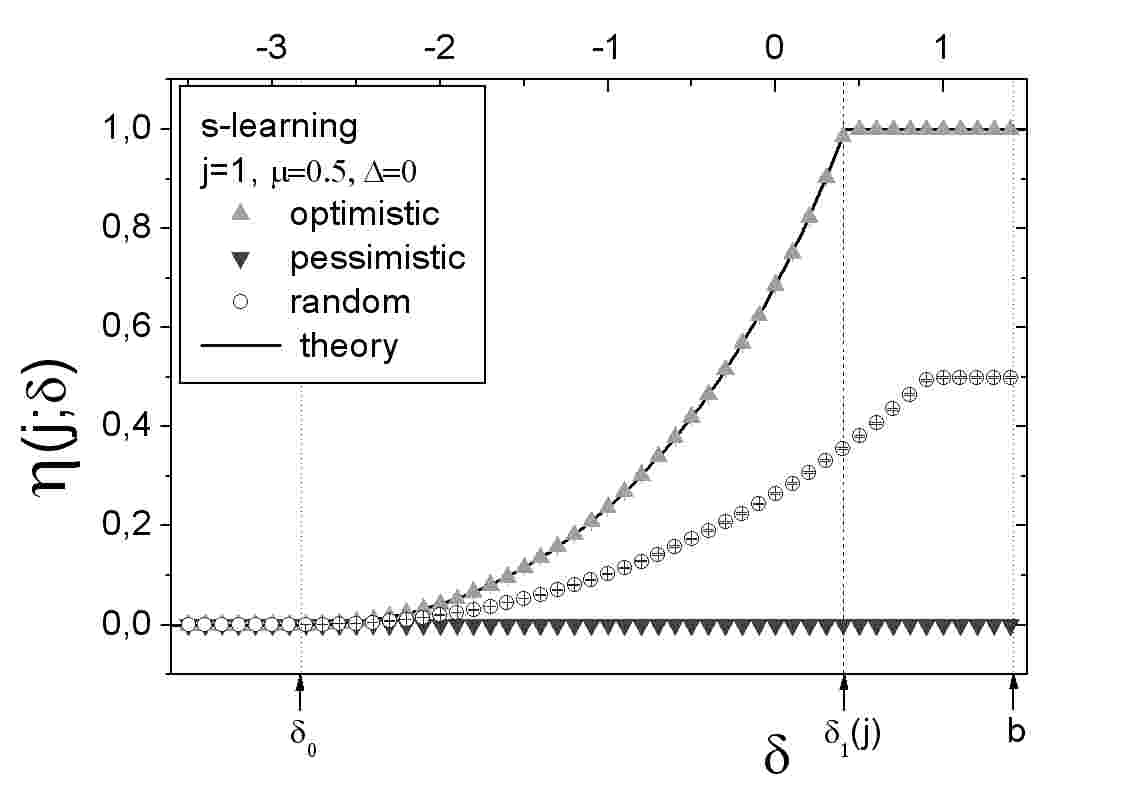} 
\includegraphics[width=7cm]{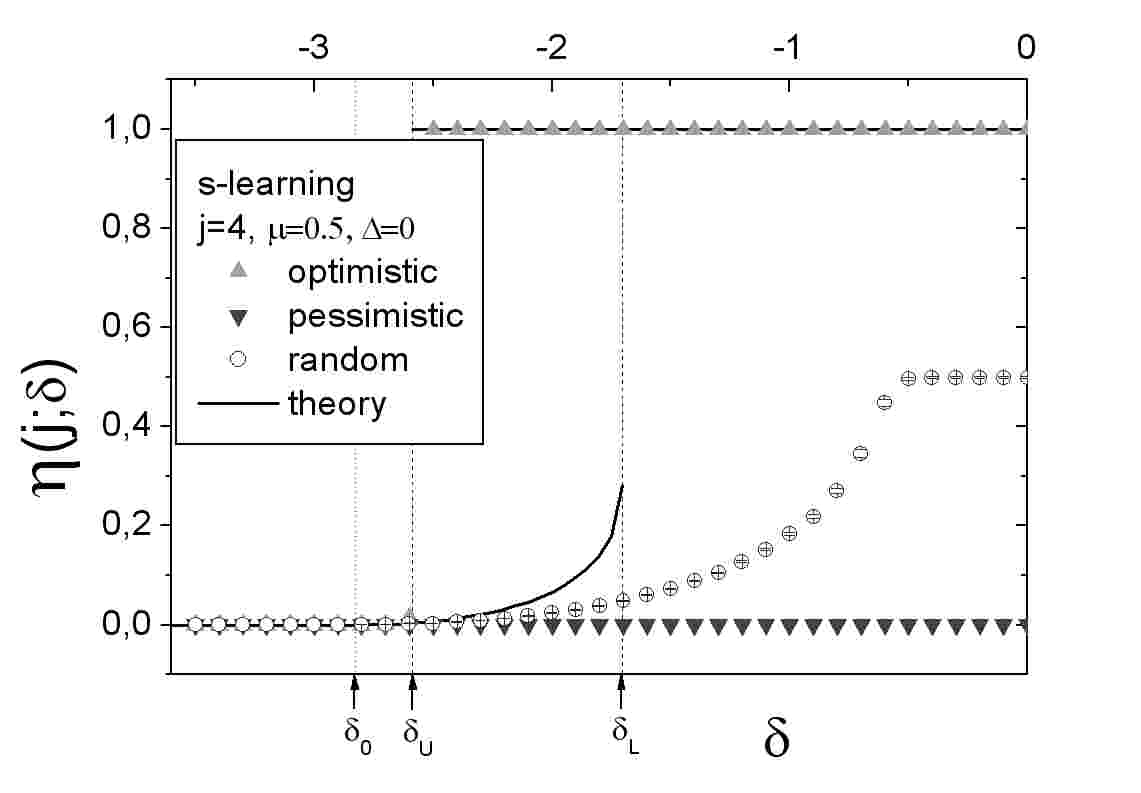} \\
\includegraphics[width=7cm]{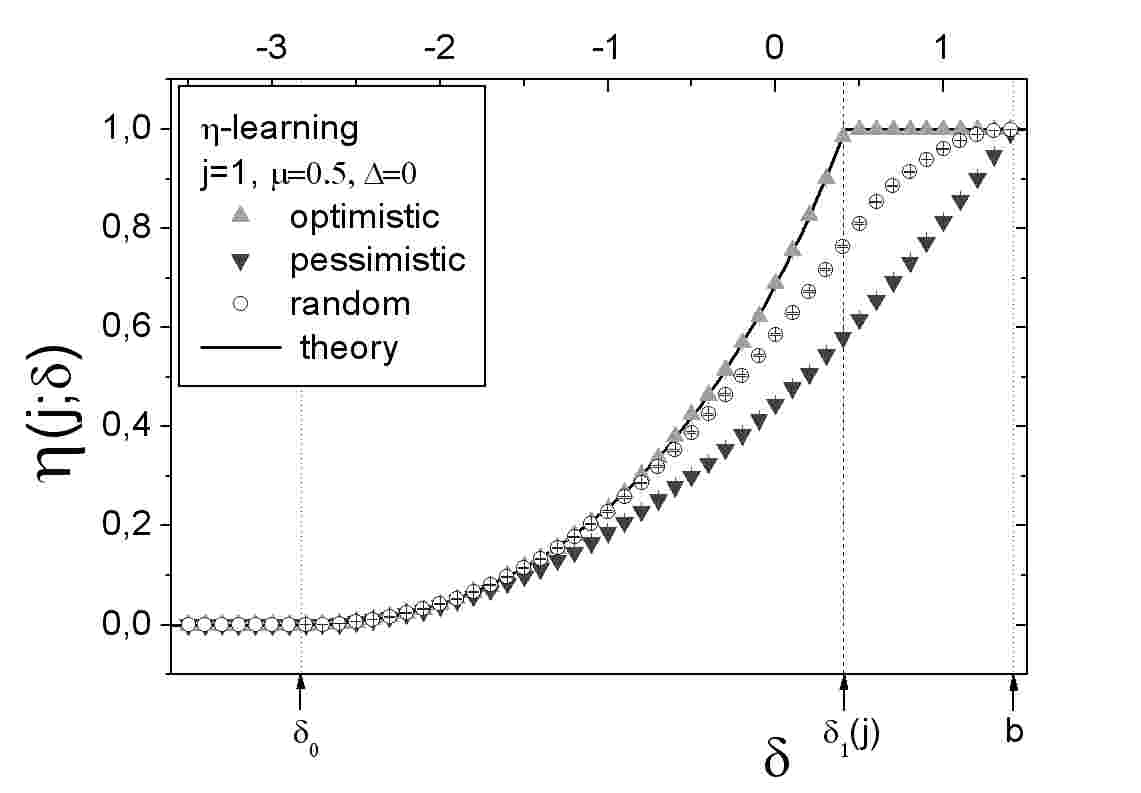} 
\includegraphics[width=7cm]{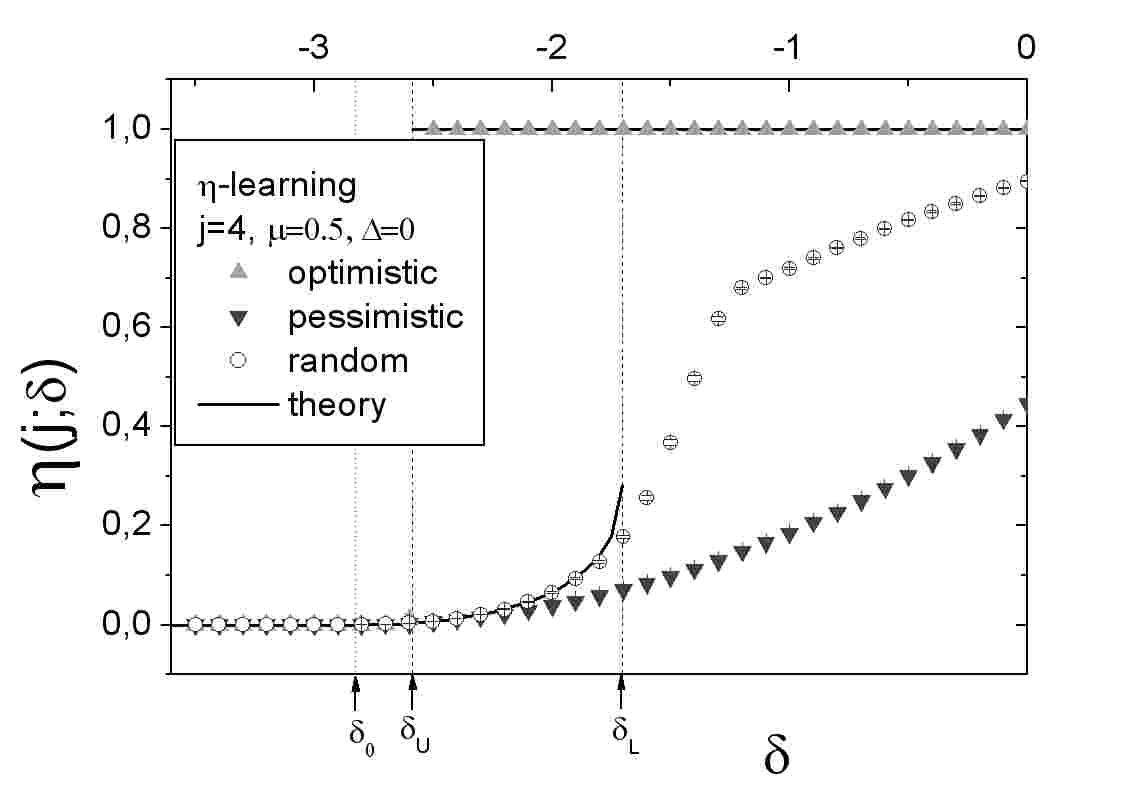} 
\end{tabular}
\caption[]{ \scriptsize{Reinforcement learning ($\mu=0.5$, $\Delta=0$). 
$\eta$ at equilibrium versus $\delta\equiv h-p$ for $j=1$ and $j=4$, for 
$s$-learning (above) and $\eta$-learning (below).}}
\label{fig.RL_eta}
\end{figure}

Like in weighted belief learning, the system's behavior with reinforcement 
learning strongly depends on the initial states. In fact, only individuals 
with $a_i(0)>0$ can actually learn from experience because their first 
decision is to buy. Therefore, the attractions of buyers (but only these) 
converge to the actual payoffs $s_i=x_i+\delta+j\eta$, both with $\eta$- 
and $s$-learning. Their values of $a_i$ at convergence present a slope $1$ 
as a function of $x_i$. With $s$-learning non-buyers (whose attractions are 
negative) cannot use the information carried by the forgone payoffs. These 
agents decrease iteratively by a factor $1-\mu$ the absolute values of 
$a_i$ at each step of the learning process. Attractions keep thus their 
negative signs: the corresponding individuals persist in state $\omega_i=0$
and the attractions of non-buyers converge to $a_i=0$ whatsoever the value 
of $x_i$. In figure \ref{fig.RL_a_iwp} the corresponding $a_i$ vs $x_i$ 
present a vanishing slope, and there are individuals with $a_i=0$ evenly 
distributed over the $x_i$ axis. Notice that even when $\delta$ is large 
enough (low enough price) to allow everybody get positive payoffs, at 
equilibrium there remain non-buyers with vanishing attractions 
independently of their value of $x_i$. On figure \ref{fig.RL_eta}, the 
fraction of buyers $\eta$ (with random initialization) is seen to be 
systematically smaller than the fraction expected from the phase diagram. 
Since in our random initialization setting  the initial values $a_i(0)$ 
are selected with equal probabilities of being positive or negative, the 
initial fraction of buyers is $\eta(0)=0.5$. Since those who begin with 
$a_i(0)<0$ are unable to change their mind, the upper bound to $\eta$ is 
$1/2$, as is seen in the upper figures \ref{fig.RL_eta}. For the same 
reasons, with the pessimistic initialization nobody buys independently of 
$\delta$. Only with the optimistic initialization all the individuals can 
learn and make correct estimations of their payoffs: the corresponding 
curves $\eta$ vs. $\delta$ are similar to those with myopic best response. 

With $\eta$-learning the behavior is closer to that of weighted belief 
learning: the curves $\eta(j;\delta)$ follow the same trends as in figures 
\ref{fig.WBL_aver_eta_d}. The attractions of buyers (non-buyers) converge 
to $a_i=\delta+x_i+j\eta$ ($a_i=\delta+x_i$). The range of $x_i$ values 
where individuals may have different $a_i$ even if they have similar $x_i$ 
is obtained like with weighted belief learning, putting $\Delta=0$ in the 
equations. This gives $-\delta-j\eta<x_i<-\delta$. This is illustrated on 
figures  \ref{fig.RL_a_iwp} (below), and explains why here also the 
values of $\eta$ at convergence are systematically smaller (or equal) to 
those expected at the Nash equilibrium, as may be seen on figures \ref
{fig.RL_eta} (below). Notice that with $\eta$-learning, even if 
non-buyers do not use the information about $\eta$, they may still buy 
provided that $x_i+\delta>0$, independently of the initial guess 
$\eta_i(0)$. This is why $\eta$ may be larger than with $s$-learning at 
convergence, see figure \ref{fig.RL_eta} (below), and even reach 
saturation provided $\delta$ is large enough. The same argument as in the 
preceding section shows that saturation can be reached only if $\delta > b$.

\section{Discussion and Conclusion}
\label{sec:conclu}
It is interesting to compare the results obtained with weighted belief 
learning to those with reinforcement learning. In both cases, the 
equilibrium values of the attractions may be calculated by replacing 
$a_i(t)$, $\eta(t)$, $\omega_i(t)$ in (\ref{eq.s_learning}) and (\ref
{eq.eta_learning}) by their asymptotic values $a_i$, $\eta$ and 
$\omega_i$. With $s$-learning these are 
$a_i=[\Delta+(1-\Delta)\omega_i]s_i$: the attractions of buyers converge 
to $a_i=s_i$, i.e. they estimate correctly their expected surplus. 
Non-buyers estimate $a_i=\Delta s_i$. With $\eta$-learning we have 
$a_i=x_i+\delta+j\eta[\Delta+(1-\Delta)\omega_i]$. Thus buyers also 
correctly estimate their expected surplus, and non-buyers underestimate 
it, since their attractions converge to $a_i=x_i+\delta+j\Delta\eta$.

With weighted belief $s$-learning, the agents always estimate the 
right sign of the attraction independently on whether they are buyers or 
non-buyers, so that the system converges to the theoretical Nash 
equilibrium despite the incorrect estimations by non-buyers. This is not 
true for reinforcement learning ($\Delta=0$), because in this case the 
attractions of non-buyers converge to $a_i = s_i \Delta=0$. As a result, 
at equilibrium we expect fewer buyers than with weighted belief learning, 
because with reinforcement learning individuals that have initial negative 
attractions persist in non-buying even if they could obtain positive 
payoffs.

With $\eta$-learning, like with $s$-learning, buyers' asymptotic 
attractions converge to the actual surpluses both with weighted belief and 
with reinforcement learning: $a_i=s_i$. Non-buyers' surplus estimations 
converges to $a_i=s_i -(1-\Delta)j\eta$, which may be negative even if 
$s_i>0$. Therefore, in contrast with $s$-learning, weighted belief $\eta$
-learning may fail to reach the theoretical Nash equilibria. With 
reinforcement learning, on the other hand, $\eta$-learning may be more 
performant than $s$-learning, since non-buyers' estimations converge to 
$a_i=\delta+x_i$, i.e., they disregard the social component of the surplus 
but take into account correctly their idiosyncratic preferences. 
Therefore, the fraction of buyers increases with $\delta$, without getting 
stuck at a value determined only by the initial conditions, as happens 
with $s$-learning.

The comparison of $s$- and $\eta$-learning with the same parameters shows 
that with weighted belief learning, $\eta$-learning converges to fewer 
buyers than $s$-learning, because in the latter case the sign of the 
surplus is correctly estimated. This is a rather counterintuitive result, 
since individuals using $s$-learning have a poorer knowledge of the payoff 
structure.  On the other hand, $\eta$-learning allows to get closer to the 
theoretical Nash equilibria because the agents know their preferences, and 
only misestimate the fraction of buyers. To summarize, with reinforcement 
learning the quality of the equilibria with the two learning scenarios is 
inversed with respect to the one in weighted belief learning. In $\eta$
-learning, agents with the \emph{a priori} knowledge about $x_i$ and $j$ 
drive the system through learning to states with larger fractions $\eta$  
than with $s$-learning, where agents do not have this \emph{a priori} 
information. 

We only considered $0\leq\Delta \leq 1$, implying that non-buyers may only 
{\em underestimate} the learned quantity (be it the forgone 
payoff or the fraction of buyers). Values $\Delta>1$ allow to modelize the 
non-buyers regret about their chosen strategy. These $\Delta$ values can 
only lead to overestimations of the learned term, helping non-buyers to 
increase faster their attractions for buying. The result would be an 
acceleration of convergence. Since buyers make correct estimations, we 
expect that, except for reinforcement $s$-learning, the final states be the 
same as with fictitious play. With reinforcement $s$-learning, the results 
would be the same as those presented here.

To conclude, our results show that systems with interacting rational 
agents with limited information may not reach the theoretical Nash 
equilibria, even when these are unique. If the social interactions are so 
strong that there are multiple Nash equilibria, the resulting collective 
state is very sensitive to the agents' initial guesses of the opportunity 
of buying. 

We restricted our simulations to systems where all the agents use the same 
learning rule. Further investigations should consider mixtures of 
different kinds of learners. 

Our agents used deterministic learning rules. One drawback is that their 
decisions are independent of the magnitude of the attraction: only its 
sign matters. Probabilistic decision rules, where the uncertainty of the 
choice is larger the closer the attraction to $0$, have been studied in a 
related model where adaptive customers have to choose 
between different sellers \cite{WeisbuchKirman,NaWeChKi}, 
in a particular context where fictitious play is not possible. 
There, the existence of multi-equilibria is shown to lead to 
a transition between an unfaithful and a faithful behaviour
(customers going to different sellers in the first case, and preferring 
one particular seller in the other case).
Within our general framework we have studied the adaptive dynamics 
with probabilistic decision rules. A typical result is that the population 
reaches states in which decisions fluctuate close to the average ones. This 
stationary regime is in general close to the `quantal response 
equilibrium' \cite{McKelveyPalfrey95} described in economics. In addition, 
a more complex stationary state can be obtained when the choice uncertainty 
is strong enough. A detailed analysis of the collective behaviour under 
such probabilistic decision rules will be presented elsewhere \cite
{NaGoSe07}.

\section*{Acknowledgements}
This work is part of the project ``ELICCIR" supported by
the joint program ``Complex Systems in Human and Social Sciences" of the 
French Ministry of Research and of the CNRS. M.B.G. and J.-P. N. are CNRS 
members. 

\bibliographystyle{plain}
\bibliography{ecobibAvsE}

\begin{thebibliography}{10}

\bibitem{Andrecut_Ali_01}
M.~Andrecut and M.~K. Ali.
\newblock Q learning in the minority game.
\newblock {\em Physical Review E}, {\bf 64}:067103, (2001).

\bibitem{Arthur94}
W.~B. Arthur.
\newblock El farol.
\newblock {\em Amer. Econ. Review}, {\bf 84}:406, (1994).

\bibitem{Becker91}
G.~S. Becker.
\newblock A note on restaurant pricing and other examples of social influences
  on price.
\newblock {\em The Journal of Political Economy}, {\bf 99}:1109--1116, (1991).

\bibitem{Benaim_Hirsch_99}
Michel Benaim and Morris~W Hirsch.
\newblock Learning processes, mixed equilibria and dynamical systems arising
  from fictitious play in perturbed games.
\newblock {\em Games and Economic Behavior}, {\bf 29}:36--72, 1999.

\bibitem{Blume93}
L.~E. Blume.
\newblock The statistical mechanics of strategic interaction.
\newblock {\em Games and Economic Behavior}, {\bf 5}:387--424, (1993).

\bibitem{BushMosteller55}
R.~Bush and F.~Mosteller.
\newblock {\em Stochastic models for learning}.
\newblock Wiley, (1955).

\bibitem{Camerer03}
C.~F. Camerer.
\newblock {\em Behavioral Game Theory}.
\newblock Princeton University Press, Princeton, New Jersey, (2003).

\bibitem{MinorityGame_book}
Damien Challet, Matteo Marsili, and Yi-Cheng Zhang.
\newblock {\em Minority Games: Interacting agents in financial markets}.
\newblock Oxford Univ Press, 2004.

\bibitem{CheungFriedman97}
Y.~W. Cheung and J.W. Friedman.
\newblock Individual learning in normal form games: Some laboratory results.
\newblock {\em Games and Economic Behavior}, {\bf 19}:46--76, (1997).

\bibitem{Cournot60}
A.~Cournot.
\newblock Recherches sur les principes mathematiques de la theorie des
  richesses.
\newblock {\em N. Bacon, Trans. [Researches in the mathematical principles of
  the theory of wealth]. London: Haffner}, (1960).

\bibitem{Durlauf97}
S.~N. Durlauf.
\newblock Statistical mechanics approaches to socioeconomic behavior.
\newblock In B.~Arthur, S.~N. Durlauf, and D.~Lane, editors, {\em The Economy
  as an Evolving Complex System II}. Santa Fe Institute Studies in the Sciences
  of Complexity, Volume XVII, Addison-Wesley Pub. Co, (1997).

\bibitem{ErevRoth98}
I.~Erev and A.~E. Roth.
\newblock Predicting how people play games: reinforcement learning in
  experimental games with unique, mixed strategy equilibria.
\newblock {\em The American Economic Review}, {\bf 88}:4:848--881, (1998).

\bibitem{Fol74}
H.~F{\"{o}}llmer.
\newblock Random economies with many interacting agents.
\newblock {\em Journal of Mathematical Economics}, {\bf 1}:1:51--62, (1974).

\bibitem{GaGeSh}
S.~Galam, Y.~Gefen, and Y.~Shapir.
\newblock Sociophysics: A mean behavior model for the process of strike.
\newblock {\em Mathematical Journal of Sociology}, {\bf 9}:1--13, (1982).

\bibitem{GlaeserScheinkman}
E.~Glaeser and J.~A. Scheinkman.
\newblock Non-market interactions.
\newblock In M.~Dewatripont, L.P. Hansen, and S.~Turnovsky, editors, {\em
  Advances in Economics and Econometrics: Theory and Applications, Eight World
  Congress}. Cambridge University Press, (2003).

\bibitem{GlaeSaceSche}
E.~L. Glaeser, B.~Sacerdote, and J.~A. Scheinkman.
\newblock Crime and social interactions.
\newblock {\em Quarterly Journal of Economics}, {\bf CXI}:507--548, (1996).

\bibitem{GoNaPhSe07}
M.~B. Gordon, J.-P. Nadal, D.~Phan, and V.~Semeshenko.
\newblock Discrete choices under social influence: generic properties.
\newblock {\em Submitted}, (2007) Working paper: {\em
  http://halshs.archives-ouvertes.fr/halshs-00135405}.

\bibitem{GoNaPhVa05}
M.~B. Gordon, J.-P. Nadal, D.~Phan, and J.~Vannimenus.
\newblock Seller's dilemma due to social interactions between customers.
\newblock {\em Physica A}, {{\bf 356}, Issues 2-4}:628--640, (2005).

\bibitem{Granovetter78}
M.~Granovetter.
\newblock Threshold models of collective behavior.
\newblock {\em American Journal of Sociology}, {\bf 83}(6):1360--1380, (1978).

\bibitem{MiKa86}
M.~Katz and C.~Shapiro.
\newblock Technology adoption in the presence of network externalities.
\newblock {\em Journal of Political Economy}, 94:822--41, (1986).

\bibitem{Kryazhimskii_etal_00}
A.~Kryazhimskii, Y.~Kaniovski, and P.~Young.
\newblock Adaptive dynamics in games played by heterogeneous populations.
\newblock {\em Games and Economic Behavior}, {\bf 31}:50--96, (2000).

\bibitem{LaslierTopolWalliser01}
J.F. Laslier, R.~Topol, and B.~Walliser.
\newblock A behaviorial learning process in games.
\newblock {\em Games and Economic Behavior}, {\bf 37}:"340--366", (2001).

\bibitem{Marsili_etal_03}
Matteo Marsili, Damien Challet, and Riccardo Zecchina.
\newblock Exact solution of a modified el farol's bar problem: Efficiency and
  the role of market impact.
\newblock {\em Physica A: Statistical Mechanics and its Applications}, {\bf
  280}, Issues 3-4:522--553, 2000, arXiv:cond-mat/9908480v3.

\bibitem{McKelveyPalfrey95}
R.~D. McKelvey and T.~R. Palfrey.
\newblock Quantal response equilibria for normal games.
\newblock {\em Games and Economic Behavior}, {\bf 7}:6--38, (1995).

\bibitem{NaGoSe07}
J.-P. Nadal, M.~B. Gordon, and V.~Semeshenko.
\newblock in preparation.

\bibitem{NaPhGoVa06}
J.-P. Nadal, D.~Phan, M.~B. Gordon, and J.~Vannimenus.
\newblock Multiple equilibria in a monopoly market with heterogeneous agents
  and externalities.
\newblock {\em Quantitative Finance}, {\bf 5}(6):557--568, (2006).

\bibitem{NaWeChKi}
J.-P. Nadal, G.~Weisbuch, O.~Chenevez, and A.~Kirman.
\newblock A formal approach to market organisation: Choice functions, mean
  field approximation and maximum entropy principle.
\newblock In J.~Lesourne and A.~Orléan, editors, {\em Advances in
  Self-Organization and Evolutionary Economics}, pages 149--159. Economica,
  London, (1998).

\bibitem{NakayamaNakamura}
Shoichiro Nakayama and Yasuyuki Nakamura.
\newblock A fashion model with social interaction.
\newblock {\em Physica A: Statistical and Theoretical Physics}, {\bf
  337}(3-4):625--634, (2004).

\bibitem{Orlean95}
A.~Orl{\'e}an.
\newblock Bayesian interactions and collective dynamics of opinion: Herd
  behaviour and mimetic contagion.
\newblock {\em Journal of Economic Behavior and Organization}, 28:257--274,
  (1995).

\bibitem{PhaSe07}
D.~Phan and V.~Semeshenko.
\newblock Equilibria in models of binary choice with heterogeneous agents and
  social influence.
\newblock {\em \emph{submitted} to European Journal of Economic and Social
  Systems}, {\bf }, (2007).

\bibitem{Rohlfs74}
J.~Rohlfs.
\newblock A theory of interdependent demand for a communications service.
\newblock {\em The Bell Journal of Economics and Management Science}, {\bf 5}
  (1):16--37, (1974).

\bibitem{Rohlfs01}
J.~Rohlfs.
\newblock {\em Bandwagon Effects in High Technology Industries}.
\newblock MIT Press, (2001).

\bibitem{SarinVahid01}
R.~Sarin and F.~Vahid.
\newblock Predicting how people play games: a simple dynamic model of choice.
\newblock {\em Games and Economic Behavior}, 34:104--122, (2001).

\bibitem{Schelling71}
T.~S. Schelling.
\newblock Dynamic models of segregation.
\newblock {\em Journal of Mathematical Sociology}, 1:143--186, (1971).

\bibitem{SeGoNaPh06}
V.~Semeshenko, M.~B. Gordon, J.-P. Nadal, and D.~Phan.
\newblock Choice under social influence: effects of learning behaviors on the
  collective dynamics.
\newblock {\em Book Chapter in Cognitive Economics: New Trends}, {\bf
  280}:{177--203}, (2006).

\bibitem{Schelling}
Schelling T.S.
\newblock {\em Micromotives and Macrobehavior}.
\newblock W.W. Norton and Co, N.LY., (1978).

\bibitem{WeisbuchKirman}
G.~Weisbuch, A.~Kirman, and D.~Herreiner.
\newblock Market organisation and trading relationships.
\newblock {\em Working paper 1996, published in: The Economic Journal}, Volume
  {\bf 110} Issue 463:411--462, (2000).

\bibitem{Young02}
H.~P. Young.
\newblock Bounded rationality and learning. on the limits of rational learning.
\newblock {\em European Economic Review}, {\bf 46}:791--799, (2002).

\end{thebibliography}
\end{document}